\documentclass[nohyper,notoc]{JHEP}

\usepackage{amsmath,euscript,array,amssymb,cite} 
\input epsf

\newcommand{\startappendix}{
\setcounter{section}{0}
\renewcommand{\thesection}{\Alph{section}}}
\newcommand{\Appendix}[1]{
\refstepcounter{section}
\begin{flushleft}
{\large\bf Appendix \thesection: #1}
\end{flushleft}}

\def\N{{\cal N}}

\def\ttau{{\tilde\tau}}

\def\Tr{{\rm Tr}}
\def\sst{\scriptscriptstyle}

\def\SU{\text{SU}}
\def\U{\text{U}}
\def\GL{\text{GL}}
\def\SL{\text{SL}}
\newcommand{\BN}{\boldsymbol{N}}

\def\Dbarslash{\,\,{\raise.15ex\hbox{/}\mkern-12mu {\bar\D}}}
\def\Dslash{\,\,{\raise.15ex\hbox{/}\mkern-12mu \D}}
\def\delslash{\,\,{\raise.15ex\hbox{/}\mkern-9mu \partial}}
\def\delbarslash{\,\,{\raise.15ex\hbox{/}\mkern-9mu {\bar\partial}}}

\def\ms{{\mathfrak M}}
\def\ns{{\mathfrak N}}

\def\S{{\EuScript S}}
\def\P{{\EuScript P}}
\def\A{{\EuScript A}}

\newcommand{\MAT}[1]{\begin{pmatrix} #1\end{pmatrix}}
\newcommand{\EQ}[1]{\begin{equation} #1 \end{equation}}
\newcommand{\AL}[1]{\begin{subequations}\begin{align} #1
\end{align}\end{subequations}}
\newcommand{\SP}[1]{\begin{equation}\begin{split} #1 \end{split}\end{equation}}

\def\np#1#2#3{{\it Nucl. Phys.} {\bf B#1} (#2) #3}

\def\prl#1#2#3{{\it Phys. Rev. Lett.} {\bf #1} (#2) #3}

\title{An Exact Elliptic Superpotential for ${\cal N}=1^*$ Deformations
of Finite $\N=2$ Gauge Theories}

\author{Nick~Dorey, Timothy J.~Hollowood
and S.~Prem~Kumar\\
Department of Physics, University of Wales Swansea,
Swansea, SA2 8PP, UK\\
E-mail: {\tt n.dorey@swan.ac.uk}, {\tt t.hollowood@swan.ac.uk},
{\tt s.p.kumar@swan.ac.uk}}

\abstract{We study relevant deformations of the ${\cal N}=2$
superconformal theory
on the world-volume of
$N$ D3 branes at an $A_{k-1}$ singularity. In particular,
we determine the vacuum structure of the
mass-deformed theory with ${\cal N}=1$ supersymmetry and show how the
different vacua are permuted by an extended duality symmetry. We then
obtain exact, modular covariant formulae (for all $k$, $N$ and
arbitrary gauge couplings) for the holomorphic  
observables in the massive vacua in two 
different ways: by  
lifting to M-theory, and by compactification to three dimensions
and subsequent use of mirror symmetry. In the latter case, we find an exact
superpotential for the model which coincides with a certain
combination of the quadratic Hamiltonians
of the spin generalization of the elliptic Calogero-Moser integrable
system.
}


\preprint{{\tt hep-th/0108221}\\SWAT-314}

\begin{document}

\section{Introduction and summary}

Conformal field theories (CFTs) play a key role in quantum field theory as
fixed points of the renormalization group which classify possible
universality classes. They can also give rise to massive theories of direct
phenomenological interest after deformation by relevant operators.
In general, little is known about the dynamics of
interacting CFTs in four dimensions. Superconformal field theories (SCFTs)
provide an important exception to this.
In particular, strong evidence for several remarkable
properties of these theories has emerged. Firstly these models
often have marginal operators which give rise to
continuous families of superconformal field theories. They also exhibit
exact dualities which act on the
marginal parameters relating regions of strong and weak coupling.
Finally, these theories have a dual description in terms of superstring theory
on a ten-dimensional space whose non-compact part is $AdS_{5}$
\cite{revads}.
The example which is best understood is ${\cal N}=4$
supersymmetric Yang-Mills (SYM) theory with gauge group $\U(N)$ or $\SU(N)$.
This theory describes the world-volume dynamics of $N$ D3 branes in Type IIB
string theory. The model has a single marginal
dimensionless coupling, $\tau=4\pi i/g^{2} + \theta/2\pi$,
formed from the gauge coupling $g$ and the vacuum angle $\theta$. In the
D-brane realization, this coincides with the complexified IIB
string coupling. The theory has an exact $\SL(2,{\mathbb Z})$
duality which acts by
modular transformations of $\tau$ and is inherited from the $S$-duality of the
IIB theory. Finally, at large $N$, the $\SU(N)$
theory has a dual description in terms of weakly-coupled IIB
superstring theory on the near horizon geometry of the D3 branes which is
$AdS_{5}\times S^{5}$.

In this paper, we will consider
a more general class of superconformal field theories with
${\cal N}=2$ supersymmetry known as $A_{k-1}$ quiver theories.
These theories arise in IIB string theory when $N$
D3 branes are placed at an elliptic $A_{k-1}$ singularity in spacetime
\cite{dougmoore}. The ${\cal N}=4$ theory is naturally thought of as the
$k=1$ member of this family. We will focus on relevant
mass-deformations which
preserve either $\N=2$ or
${\cal N}=1$ supersymmetry. Following the convention
of the ${\cal N}=4$ case, we will refer to these as the ${\cal N}=2^{*}$
and ${\cal N}=1^{*}$ quiver theories respectively. Our main results, which
generalize previous work in the $k=1$ case \cite{donwitt,Nick},
concern the vacuum
structure and holomorphic observables of the ${\cal N}=1^{*}$ quiver theories.
We study the massive vacua of these theories using two different methods.
Firstly, we analyse the maximal degenerations of the complex curve
$\Sigma$ which
governs the Coulomb branch of the corresponding
${\cal N}=2^{*}$ theory \cite{wittm}. Secondly, we
propose an exact superpotential for the theory which coincides with the
Hamiltonian of a certain classical integrable system known as 
the {\it elliptic spin Calogero-Moser system\/} 
\cite{GH,Krichever:1994vg,Nekrasov:1996nq}.
The ${\cal N}=1^{*}$ vacua can
then be analysed directly by stationarizing this superpotential.
The two approaches are shown to be in complete agreement.
In the rest of this introduction we will outline our results. The detailed
calculations are provided in subsequent sections.

The $A_{k-1}$ quiver theory has gauge group
$G=\U(1)\times\SU(N)^{k}$ with matter in bi-fundamental representations as
determined by a quiver construction based on the Dynkin diagram of the
Lie group $A_{k-1}$.
The matter content ensures that the beta functions of
each of the $k$ non-abelian gauge couplings $g_{i}$, $i=1,\ldots,k$
vanish exactly and that the theory therefore has $k$ exactly
marginal complex couplings
\EQ{
\tau_i={4\pi i\over
g^2_i}+{\theta_i\over2\pi}\ ,
\label{couplings}
}
$i=1,\ldots,k$, where $\theta_{i}$ are the vacuum angles of each
$\SU(N)$ factor in $G$. As mentioned above, these theories arise on the
world-volume of $N$ D3 branes placed at an elliptic $A_{k-1}$ singularity
in spacetime.
In this construction, the IIB string coupling $\tau$ is
identified with the coupling
$\sum_{i=1}^{k} \tau_{i}$ of the diagonal $\SU(N)$ subgroup of $G$.
An alternative IIA brane configuration of $N$ D4-branes intersecting $k$
NS5 branes wrapped on a circle arises after T-duality in the compact
direction of the elliptic singularity. As we review below, the
strong coupling properties of the theory can then be analysed, as in
\cite{wittm},  by lifting the
IIA branes to M-theory five-branes in eleven dimensions. The M-theory
spacetime includes a torus of complex structure $\tau$ with $k$ marked
points whose relative positions on the torus encode the remaining
$k-1$ independent gauge couplings.
In the ${\cal N}=4$ case ($k=1$) modular transformations of the spacetime
torus yield the familiar
geometrical realization of IIB $S$-duality in M-theory.
For $k>1$, the theory has an enlarged $S$-duality
group corresponding to the modular group of a torus with $k$ marked
points which encode the positions of the NS5's.
In addition to modular transformations of
$\tau=\sum_{i=1}^{k} \tau_{i}$, this includes shifts of the individual
marked points by periods of the torus, physically realized as
the movement of
NS5-branes around non-trivial cycles, and other non-trivial
dualities. The IIB set-up of branes at an orbifold singularity also yields
a large-$N$ closed string dual of the quiver theories.
In this case the relevant near-horizon geometry is the
orbifold $AdS_{5}\times S^{5}/{\mathbb Z}_{k}$ \cite{kachsilv}.

As usual, a useful way to study a conformal theory is via its relevant
perturbations. The only such perturbations which preserve the full
${\cal N}=2$ supersymmetry are mass-terms for the $k$ bi-fundamental
hypermultiplets. After the deformation the resulting ${\cal N}=2$ theory
has a non-trivial Coulomb branch which can be studied by the methods of
Seiberg and Witten \cite{swone}.
As usual the Coulomb branch of the $A_{k-1}$ theory can be described as the
moduli-space of a complex curve $\Sigma$ whose genus is equal to the
rank of the gauge group, which is $r=k(N-1)+1$.
The periods of a certain holomorphic one-form on the curve
then yield the exact mass formula for BPS states of the theory.
In fact, hypermultiplet masses are easily included in the IIA brane
construction described above and lifting the IIA branes
to M-theory yields a single M5-brane wrapped on $\Sigma\times{\mathbb R}^{4}$
\cite{wittm}. This provides an explicit
construction of $\Sigma$ as an $N$-fold cover of the torus in
spacetime with prescribed singularities at the $k$ marked points. The
action of the extended $S$-duality group described above is manifest in this
approach.

As for any ${\cal N}=2$ theory in four dimensions,
the resulting Coulomb branch is a special K\"{a}hler
manifold of complex dimension $r$, with singular submanifolds where
BPS states become massless. For generic values of the masses, there are
isolated points on the Coulomb branch at which the maximal number of
mutually-local BPS states become massless. These points are special
because they survive soft-breaking of ${\cal N}=2$ SUSY down to
${\cal N}=1$ and yield ${\cal N}=1$ vacua where the theory is realized
in different massive phases.
Most famously, there are points on the Coulomb branch where magnetic
monopoles become massless and, after soft-breaking, condense in the vacuum
leading to the confinement of electric charges. When combined with the known
holomorphy properties, this also provides a powerful approach for obtaining
exact results for the vacuum properties of ${\cal N}=1$ theories. In this
paper we will present an exact analysis of the maximally singular points on
the Coulomb branch of the $A_{k-1}$ quiver theories for all $k$ and $N$ and
the resulting massive ${\cal N}=1$ vacua. These results generalize
the existing knowledge about massive deformations of the ${\cal N}=4$
theory \cite{donwitt,Nick} which we will now briefly review.
In terms of ${\cal N}=2$ supersymmetry, the ${\cal N}=4$ theory contains a
vector multiplet and a single hypermultiplet in the adjoint representation
of the gauge group. The most general relevant deformation which preserves
${\cal N}=2$ supersymmetry consists of introducing a mass term for the
adjoint hypermultiplet. Following \cite{polstr}, the resulting theory will be
referred to as ${\cal N}=2^{*}$ SUSY Yang-Mills.
The Coulomb branch of the ${\cal N}=2^{*}$ theory
with gauge group $\SU(N)$ 
is described by a branched $N$-fold
cover of the standard flat torus $E_\tau$
with complex structure parameter $\tau$ which is part of the
spacetime of the M-theory brane construction. The
points on the Coulomb branch which yield massive ${\cal N}=1^{*}$ vacua
correspond to the maximal degenerations of the curve. These are
{\it unbranched} (unramified)
$N$-fold covers of the torus $E_\tau$ \cite{donwitt},
which are themselves tori with complex structure parameter $\tilde{\tau}$
of the form
\EQ{
\tilde{\tau}=\frac{q\tau+l}{p}\qquad
\text{with}\qquad p\,q=N\quad\text{ and }\quad l=0,\ldots,p-1\ .
}
The total number of massive ${\cal N}=1^{*}$
vacua is therefore equal to $\sum_{p|N}p$. As each of these vacua is
associated with the condensation of a BPS state with definite electric and
magnetic charges, the theory in each vacuum is not invariant under $S$-duality.
Instead, $S$-duality permutes the massive ${\cal N}=1$ vacua,
relating the physics of the
theory in one ground-state at one value of the coupling to that of the theory
in another ground-state at a different value of the coupling. Hence,
one might say that the $S$-duality of the underlying
${\cal N}=4$ theory is ``spontaneously broken'' in the
${\cal N}=1^{*}$ theory.  The action of
$\SL(2,{\mathbb Z})$ on the vacua is the
same as the natural action of $\SL(2,{\mathbb Z})$ on the
set of $N$-fold covers of the torus $E_\tau$ \cite{donwitt}.

Despite the ``spontaneous breaking''
of $S$-duality described above, the theory in
each massive ${\cal N}=1^{*}$ vacuum has a novel kind of duality named
$\tilde{S}$-duality in \cite{oferandus}.
This duality reflects the fact that the degenerate
curve describing each massive vacuum is a torus with complex structure
parameter $\tilde{\tau}$ given above. The low-energy physics of the
${\cal N}=2^{*}$ theory (and its ${\cal N}=1^{*}$ deformation) only depends
on the complex structure of the curve and therefore is invariant
under modular transformations acting on $\tilde{\tau}$. $\tilde{S}$-duality
therefore has the same status as the IR electric-magnetic duality of
Seiberg-Witten theory and is not expected to be valid at all length-scales.
Nevertheless, this new duality leads to interesting predictions for the
behaviour of the theory in the limit of large 't Hooft coupling which
can be compared directly with the IIB dual background studied by
Polchinski and Strassler \cite{polstr}.

One of the main aims of this paper is to extend our understanding of
supersymmetry-preserving relevant deformations to the $A_{k-1}$ quiver
models. In fact we find natural generalizations of the phenomena described
above which are familiar from the ${\cal N}=4$ case. Specifically, we study
the maximally singular curves of the ${\cal N}=2^{*}$ quiver theories
which are again unbranched $N$-fold covers of the spacetime torus $E_\tau$.
We find:

{\bf 1.} The extended $S$-duality group described above is
``spontaneously broken''
and has a non-trivial action on the set of massive ${\cal N}=1^{*}$ vacua.
In particular, shifts of the $k$ individual marked points
by periods of $E_\tau$ now generate an additional degeneracy of
$N^{k-1}$ massive vacua for each unbranched $N$-fold cover $E_{\tilde\tau}$.
The total number of massive vacua is therefore $N^{k-1}\sum_{p|N} p$.

{\bf 2.} The $\tilde{S}$-duality group of each massive vacuum is extended
in the obvious way to the modular group of a torus of complex structure
parameter $\tilde{\tau}$ with $k$ marked points. In particular shifts of the
individual marked points by periods of $E_{\tilde\tau}$ are
exact dualities of the IR physics in each vacuum.

One of the most fascinating and mysterious aspects of ${\cal N}=2$
supersymmetric gauge theories in four dimensions is their relation to
finite-dimensional classical integrable systems
\cite{MM,martinec,donwitt}. In this paper,
we will provide results which extend this correspondence in
new directions and harness it to provide a practical method of computing
physical quantities. We will now review the connection between
${\cal N}=2$ SUSY and classical integrability in the context of the
mass-deformed ${\cal N}=4$ theory and briefly summarize our new results.

As discussed above, the ${\cal N}=2^{*}$ theory with gauge group $\U(N)$
has a Coulomb branch of complex dimension $N$ which is the moduli-space of
a Riemann surface $\Sigma$ of genus $N$. In \cite{donwitt}, Donagi and
Witten gave a concrete recipe for constructing $\Sigma$ as the spectral
curve of a certain classical integrable system.
The phase space of the system in
question is the moduli space of solutions of a set of two-dimensional
field equations for a $\U(N)$ gauge field and adjoint Higgs known as
Hitchin's equations. The equations are obtained from the dimensional
reduction of the self-dual $\U(N)$ Yang-Mills equation in four dimensions
to the two dimensional torus $E_\tau$ by including certain $\delta$-function
source terms. As usual for self-duality equations, the resulting
moduli-space is a hyper-K\"{a}hler manifold. One of the
three linearly-independent complex structures of this manifold will play a
special role in the following discussion. In particular, the resulting
complex symplectic manifold has a very
concrete realization as the complexified phase space of the elliptic
Calogero-Moser model \cite{Nekrasov:1996nq}.

The Calogero-Moser model is an integrable system of
$k$ non-relativistic particles with positions
$X_{a}$ and momenta $p_{a}$, $a=1,\ldots, N$,
interacting via the pairwise potential \cite{oandp},
\begin{equation}
V= \sum_{a>b}\, \wp(X_{a}-X_{b})\ ,
\end{equation}
where $\wp(X)$ is the Weierstrass function. The system is integrable
because of the existence of $N$ Poisson-commuting Hamiltonians
$H_{a}$, $a=1,\ldots, N$. The explicit
integration of Hamilton's equations (see the Appendix of \cite{oandp})
proceeds by
identifying the appropriate set of action-angle variables with respect to
which the equations of motion become linear. While the action variables are
just the $k$ commuting Hamiltonians themselves, the corresponding angle
variables are the flat coordinates on a $k$-dimensional torus
in phase space. It is a
highly non-trivial fact that this torus is precisely
the Jacobian variety of an appropriate
Riemann surface of genus $N$, known as the
spectral curve. Changing back to the original variables, an explicit solution
for the $N$ positions $X_{a}(t)$, for any set of initial data,
can then be given in terms of $\theta$-functions on the Jacobian.
The connection to ${\cal N}=2$ supersymmetry in four
dimensions starts from the observation that the
complex curve $\Sigma$ which governs the Coulomb branch of the
${\cal N}=2^{*}$ theory with gauge group $\U(N)$ is precisely
the spectral curve of the Calogero-Moser system. In particular, the curves
coincide if
we promote the $N$ positions and momenta to complex numbers and identify
the resulting holomorphic Hamiltonians $H_{a}$ with the $N$ order parameters
$u_{a}=\langle {\rm Tr}\, \Phi^{a} \rangle$ of the ${\cal N}=2$ theory.
(Here, $\Phi$ is the adjoint scalar in the $\U(N)$ vector multiplet.)
In particular, the first non-trivial Hamiltonian
\begin{equation}
H=\sum_{a=1}^{N} \frac{p^{2}_{a}}{2} -m^2
\sum_{a\neq b}\, {\wp}(X_{a}-X_{b})
\label{ham}
\end{equation}
is identified with the quadratic order parameter $u_{2}$.

The correspondence as stated above is rather abstract as, so far we only
have an ${\cal N}=2$ interpretation for the conserved quantities and not the
corresponding angle variables. Fortunately, a much clearer picture emerges
after compactification of the four-dimensional theory to three dimensions
on a circle \cite{Kapustin:1998xn}.
Now the theory acquires new scalar degrees of freedom from the
dimensional reduction of the $N$ massless abelian gauge fields on the
four-dimensional Coulomb branch. The component of each four-dimensional
gauge field in the reduced direction gives rise
to a Wilson line while the remaining components can be dualized in three
dimensions in favour of another scalar field. As first explained by
Seiberg and Witten \cite{sw3d}, these variables naturally lie on a torus
whose periods are controlled by the effective abelian couplings of the
low-energy theory in four dimensions. More precisely, they take values
on the Jacobian variety of $\Sigma$ and can therefore be
identified with the
angle-variables of the complexified Calogero-Moser system.

After including both the $N$ complex moduli $u_{a}$ of the four-dimensional
theory and the $2N$ new periodic scalars discussed above,
the Coulomb branch, which we denote throughout as $\ms$,
of the compactified theory is a manifold of real-dimension
$4N$. The low-energy effective action is a three-dimensional non-linear
$\sigma$-model with target $\ms$. As the theory has eight supercharges
this manifold must be hyper-K\"{a}hler. In fact
the hyper-K\"{a}hler manifold in question is precisely
the phase space of the (complexified) 
classical integrable system described above.
In particular, Seiberg and Witten \cite{sw3d}
have argued that $\ms$ has a preferred
complex structure which is independent of the radius of compactification.
With respect to this complex structure, the manifold has a description as a
toric fibration where the four-dimensional Coulomb branch parameterized by the
moduli $u_{a}$ and the Jacobian of the curve $\Sigma$ is the fibre above
each point. As above, the moduli $u_{a}$ are identified with the
$N$ commuting Hamiltonians of the integrable system, while
the holomorphic coordinates on the fibre are just the corresponding
angle variables. Thus, with respect to the preferred complex structure,
the Coulomb branch of the compactified theory coincides
with the complexified phase space of the elliptic Calogero-Moser model.
In fact, one may also demonstrate directly the equivalence of $\ms$ and
the moduli space of Hitchin's equations as hyper-K\"{a}hler manifolds.
This equivalence turns out to be a generalization of the mirror
symmetry between three-dimensional gauge theories with eight supercharges
discovered in \cite{IS}.

It is also interesting to consider soft breaking to ${\cal N}=1$
supersymmetry from the point of view of the compactified theory.
As in four dimensions, this is accomplished by introducing a non-zero mass
$\mu$ for
the adjoint chiral multiplet in the ${\cal N}=2$ vector multiplet, or
in other words adding  
the perturbation $\mu u_{2}$ to the superpotential. This superpotential
lifts the Coulomb branch leaving only isolated vacua as in the uncompactified
theory. One of the main results of \cite{Nick} was that the
integrable systems viewpoint provides a simple and quantitative
description of this effect. Essentially all we need is the identification of
the Coulomb branch of the compactified theory as the phase-space of the
Calogero-Moser model with the complex positions $X_{a}$ and momenta
$p_a$ providing a convenient choice of holomorphic coordinates. In
terms of these variables, the superpotential $u_{2}$ is simply given by the
Calogero-Moser Hamiltonian given in \eqref{ham} above. This leads to
a new connection between ${\cal N}=1$ supersymmetric gauge theories and
integrable systems: the vacua of the former can be identified with the
equilibrium configurations of the latter. Numerous checks of this
identification were presented in \cite{Nick}. It also provides a
practical method of computing the condensates of chiral operators for the
${\cal N}=1^{*}$ theory \cite{nickprem,oferandus}.

In the preceding discussion, we outlined the correspondence between the
massive deformations of ${\cal N}=4$ SUSY Yang-Mills and a certain classical
integrable system. One of the main goals of the present paper is to extend
this to massive deformations of the $A_{k-1}$ quiver models for $k>1$. The
first step is to identify the appropriate integrable model. Fortunately,
several relevant results already exist. In particular,
Kapustin has used the mirror
symmetry of \cite{IS} to show that the Coulomb branch of the ${\cal N}=2^{*}$
quiver theories coincides with the moduli space of an appropriate
system of self-duality equations. For the $A_{k-1}$ theory,
the relevant equations are the $\U(N)$ Hitchin equations on the torus
$E_\tau$ with specified behaviour at $k$ punctures. In the following we will
confirm that the spectral curve of this system coincides with the
complex curve for the $A_{k-1}$ elliptic model given by Witten. However our
main interest will be to recast this system as a system of interacting
particles with ``spin'' in one dimension, analogous to the elliptic
Calogero-Moser system of the $k=1$ case. Again, the required result is
available in the
existing literature. Nekrasov has shown that the $\U(N)$ Hitchin system on a
torus with $k$ punctures can be recast as the spin generalization of the
elliptic Calogero-Moser system 
\cite{Nekrasov:1996nq}. This system
consists of $N$ particles 
moving on a circle, each carrying $k$ ``spin''
variables. The Hamiltonian is given explicitly in \eqref{spot} below.
As in the $k=1$ case, we investigate the connection between the maximally
degenerate curves, the vacua of the theory
obtained by soft-breaking of ${\cal N}=2$ supersymmetry to ${\cal N}=1$ and
the equilibrium configurations of the integrable system. In line with the
arguments presented above, our main result is that the exact superpotential
of the theory is precisely a certain linear combination of the
quadratic Hamiltonians of the elliptic spin-Calogero-Moser
system. In particular, we
show that the equilibrium configurations of the elliptic spin system
are in one-to-one correspondence with the genus-one degenerations of the
complex curve $\Sigma$. Finally we perform an explicit computation of
condensates in both approaches and demonstrate precise agreement.

The paper is organized as follows. In Section 2, we deduce the vacuum
structure of the mass-deformed quiver theory by analysing its
classical superpotential. The resulting structure has a very nice
interpretation 
in terms of brane configurations in Type IIA string theory. In
Section 3, we recall, following Witten \cite{wittm}, how these
elliptic models can be solved by lifting the Type IIA brane
configurations to M Theory. In this picture, the $\N=1^*$ deformation
can be described in terms of a rotation whose description is relegated
to Appendix B. The net result is that the massive vacua of the mass
deformed theory have a very simple interpretation in the M Theory
picture which allows us to solve for the holomorphic
observables. Section 4 describes a rather different way to probe
the vacuum structure via the compactification scenario described
above.
The net result is an expression for the exact superpotential of the
low energy three-dimensional $\sigma$-model which encodes via its
minima all the holomorphic structure of the vacua.

\section{The ${\cal N}=1^*$ deformations of the quiver theory}

We begin by recalling the matter content of the $A_{k-1}$ quiver
gauge theories.
They have $\N=2$ supersymmetry and gauge
group $\U(N)^k$ with $k$ hypermultiplets $(Q_{i,i+1},
\tilde{Q}_{i+1,i})$, $i=1,\ldots k$ (the subscripts being defined modulo
$k$) transforming in
the bi-fundamental representation $((\BN,\overline{\BN}),
(\overline{\BN},\BN))$ of
the $i$-th and $i+1$-th $\U(N)$ factors. In terms of $\N=1$
superfields we denote
the $\N=2$ vector multiplets as $\{W^\alpha_i, \Phi_i\}$;
$i=1,2,\ldots,k$. As each $\U(N)$ factor sees $2N$ flavors in the
fundamental representation, these $\N=2$ theories have vanishing
$\beta$-functions. In particular, in the absence of hypermultiplet
masses they are superconformal theories. Consequently there are $k$
finite and independent gauge couplings
\EQ{
\tau_i={4\pi i\over
g^2_i}+{\theta_i\over2\pi}\ ,
}
$i=1,\ldots,k$, each corresponding to an exactly marginal
direction in the space of quiver theories \cite{kachsilv}.
Introducing an overall gauge
coupling $\tau\equiv 4\pi i/
g^2+{\theta/2\pi}=\sum_{i=1}^k{\tau_i}$ to be thought of as the
coupling for the diagonal $\U(N)$, the theory
has a superpotential
\EQ{
W=\frac1{g^2}\sum_{i=1}^k\Tr\;\Phi_i[\Phi_{i,i+1},\Phi_{i+1,i}]\ .
}
We emphasize that the labels are defined modulo $k$.
In what follows the dynamics of the
$\U(1)$ factors of the $\U(N)$ groups is often irrelevant and can be
ignored as they decouple from the interacting theory in the infra-red.

We will consider ``$\N=1^*$
deformations'' of these theories, {\it i.e.\/}~soft breaking to $\N=1$
with generic hypermultiplet masses $m_i$ and generic masses $\mu_i$
for the adjoint chiral
multiplets via the following $\N=1$ tree-level superpotential,
\EQ{
W= {1\over {g^2}}\sum_{i=1}^k
\Tr\;\Big\{\Phi_i\big[\Phi_{i,i+1},\Phi_{i+1,i}]
-m_i\,\Phi_{i,i+1}\Phi_{i+1,i}-\mu_i\,\Phi_i^2\Big\}\ .
\label{suppottree}
}
We refer to this
theory as the ``$\N=1^*$ quiver theory''. The theory with $\mu_i=0$ and
$m_i\neq 0$ which we dub the ``$\N=2^*$ quiver theory'', will play a
central r\^ole in our analysis of the classical and quantum vacuum
structure of the $\N=1^*$ deformation.

The $A_{k-1}$ quiver theories with and without hypermultiplet masses
(also known as the elliptic models) are
obtained in Type IIA/M-theory \cite{wittm} on the world-volume
of $N$ D4-branes suspended between $k$ parallel
NS5-branes arranged on a compact direction in spacetime as in
Fig.~1.\footnote{As is well known, in the T-dual Type IIB picture, the
quiver gauge
theory may be obtained as the low-energy world volume dynamics
of $N$ D3-branes at a ${\mathbb Z}_k$ orbifold singularity in an $A_{k-1}$ ALE
space \cite{dougmoore}.}
Following the usual conventions of \cite{wittm} we choose our
chain of $k$ NS5-branes to be located at $x^7=x^8=x^9=0$ and at
specific values
$\{x^6_i\}$ of the $x^6$-coordinate. The $x^6$ direction is
compactified on a circle of radius
$L$. There are $N$ (fractional) D4-branes between the $i-1$-th and $i$-th
five-branes, with world-volumes parameterized by $x^0,x^1,x^2,x^3$ and
$x^6$. The (classical) position of a D4-brane may be conveniently
described by the complex variable $v=x^4+ix^5$.
At low energies this construction actually gives rise to a
$4$-dimensional $\SU(N)^k\times \U(1)$
theory with the desired matter content.
The missing $\U(1)$ factors
corresponding to the center-of-mass motion in the $x^4-x^5$ plane of all the
four-branes in a given segment between adjacent NS5-branes
are IR-free and decouple.\footnote{Actually this is not quite true
since the overall $\U(1)$ remains dynamical. But this decouples from
the other fields and we ignore it in what follows.} The
distance between the $i$-th and $i+1$-th five-branes is directly related
to the gauge coupling of the $i$-th $\SU(N)$ factor, $
g^{-2}_i=(x^6_{i+1}-x_{i}^6)/(8\pi^2g_sL)$ where $g_s$ is the Type IIA
string coupling. Clearly then, the overall coupling $1/g^2=\sum_i
1/g^2_i$ is proportional to the radius of the $x^6$-dimension and is also
the gauge coupling for the $\U(1)$ factor.
\begin{figure}
\begin{center}\mbox{\epsfxsize=3in\epsfbox{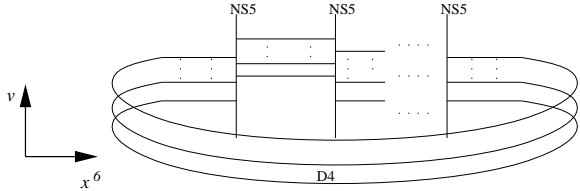}}\end{center}
\caption{\small The Type IIA setup of the elliptic model consisting of
four-branes suspended between
five-branes arranged on a compact direction.}
\end{figure}
As we have already mentioned,
the difference in positions of the centers-of-mass of
two neighbouring stacks in Fig.~1 is frozen in the infra-red and is
therefore a modulus which is identified with the
the hypermultiplet
mass-parameter $m_i$ in the low energy theory.
\begin{figure}
\begin{center}
\mbox{\epsfxsize=1.5in\epsfbox{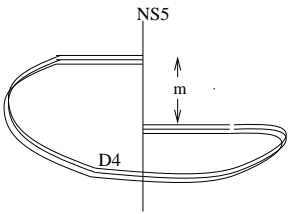}}
\end{center}
\caption{\small $\N=2^*$ SUSY Yang-Mills realized in Type IIA theory.}
\end{figure}

The periodicity of the
spacetime implies that
there are only $k-1$ independent masses with $\sum_{i=1}^k m_i=0$. As
explained in \cite{wittm} this restriction may be avoided by modifying the
M-theory spacetime so that upon going around the $x^6$ circle the
value of $v$ is shifted at some point by an amount $m$ leading to
\EQ{
\sum_{i=1}^km_i=m\ .
\label{gmass}
}
The parameter $m$ is often referred to as the ``global mass''. The
position of the global shift is completely unphysical and can be
chosen anywhere.
The simplest such setup (Fig.~2) with a single five-brane and
$N$ D4-branes yields the $\SU(N)$, $\N=2$ theory with a massive adjoint
hypermultiplet---the so-called $\N=2^*$ theory, first analyzed by
Donagi and Witten \cite{donwitt}.

Soft breaking of the above set-up to the $\N=1^*$ quiver theory
is achieved by introducing generic masses $\mu_i$ for the adjoint
chiral multiplets $\Phi_i$. This can be
realized in the brane setup by rotating the NS5-branes relative to
each other in the $(v,w)$-plane $(w=x^8+ix^9)$. This feature is
described separately in Appendix B where we generalize the approach of
\cite{Hori:1998ab} to the elliptic models.

\subsection{Ground states of the $\N=1^*$ quiver theory}

In this section, we determine the vacuum states of the $\N=1^*$ quiver
theories generalizing the structure in the basic $\N=1^*$ case
\cite{donwitt,vafawitt}. It is worth reviewing the latter in some
detail because this will provide us with the necessary intuition for
the quiver generalization.

In the mass deformed $\N=4$ theory, the vacuum structure can be
uncovered by first establishing a classical picture which can then be
rather easily refined to give the full quantum description.
The classical vacua are governed by the tree-level
superpotential \eqref{suppottree}
for $k=1$. As usual in an $\N=1$ theory, we
can avoid solving the $D$-flatness conditions by solving the $F$-flatness
conditions modulo complex gauge transformations. In the $k=1$ case,
these are
\AL{
&[\Phi,\Phi^\pm]=m\Phi^\pm\ ,\label{wwa}\\
&[\Phi^+,\Phi^-]=2\mu\Phi\ .\label{wwb}
}
modulo $\SL(N,{\mathbb C})$. The independent solutions are associated to the
partitions $N=q_1+\cdots+q_p$ with
\EQ{
\Phi_c=M_c\MAT{J_c^{(q_1)}&&&\\ &J_c^{(q_2)}&&\\ &&\ddots&\\
&&&J_c^{(q_p)}}\ ,
\label{ggs}
}
where $c=1,2,3$, $\Phi\equiv\Phi_3$, $\Phi^\pm\equiv\Phi_1\pm i\Phi_2$
and $J_c^{(q)}$ are the generators of the $q$-dimensional
representation of $\SU(2)$. The mass parameters are
$M_1=M_2=\sqrt{2\mu m}$
and $M_3=m$. The solution generically leaves abelian
factors of the gauge group unbroken and the vacuum is
massless. However, for equipartitions, $N=q\cdot p$, the unbroken
gauge symmetry is $\SU(p)$. In these cases,
the non-abelian symmetry leads to confinement in the infra-red with a
mass gap. The number of quantum vacua of an $\SU(p)$ gauge theory with
$\N=1$ supersymmetry is $p$ and so the total number of massive vacua
is
\EQ{
\#\big(\text{massive vacua of }\N=1^*\big)=\sum_{p|N}p\ ,
}
a sum over the divisors $p$ of $N$.

{\it The Type IIA brane picture}

Before analyzing the classical vacuum structure of the $\N=1^*$
quiver theory,
it will be extremely instructive to see how the standard enumeration
of massive vacua of $\N=1^*$ theory (mass-deformed $\N=4$ theory)
described above 
arises from the viewpoint of the Type IIA brane picture. In
particular, the massive vacua are associated to
special (classical) brane configurations corresponding
to maximally singular points on the Coulomb branch of the $\N=2^*$ theory,
where new massless (string) states appear.
Importantly, the massless
string states should give rise to charged hypermultiplets for all the
classically visible $\U(1)$ factors at the aforementioned Coulomb
branch singularities.
This ensures that upon further soft breaking to $\N=1^*$ all these $\U(1)$
factors get Higgsed and only a non-Abelian gauge symmetry survives
and a mass gap is generated.
It is easy to see that these requirements lead us to
configurations where the branes are arranged in
``helices'' as in Figs.~3(a) and 3(b). The helices arise in the
following way. With only an $\N=2^*$ mass ($\mu=0$) the $F$-flatness conditions
\eqref{wwa}-\eqref{wwb} allow a more general solution for $\Phi$:
\EQ{
\Phi=m\MAT{c_11_{\sst[q_1]\times[q_1]}+
J_3^{(q_1)}&&&\\ &c_21_{\sst[q_2]\times[q_2]}+J_3^{(q_2)}&&\\ &&\ddots&\\
&&&c_p1_{\sst[q_p]\times[q_p]}+J_3^{(q_p)}}\ ,
\label{ggst}
}
for arbitrary constants $c_r$. Each of the $p$ blocks corresponds to a
helix and the parameters $c_r$ encode the centre-of-mass position of
each helix. Notice that the pitch of a helix as one
goes around $x^6$ is $m$, where
$m$ is the $\N=2^*$ global mass.
\begin{figure}
\begin{center}
\mbox{\epsfxsize=3.5in\epsfbox{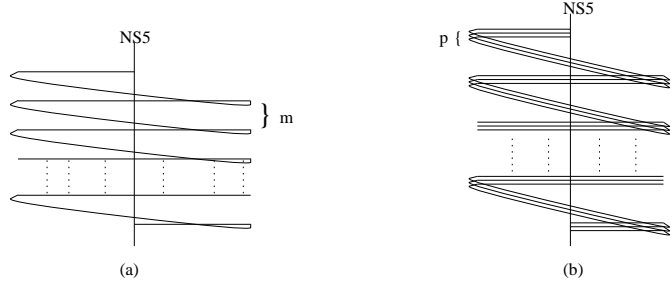}}
\end{center}
\caption{\small (a)
The Coulomb branch singularity of $\N=2^*$ SUSY Yang-Mills
which yields the Higgs vacuum of $\N=1^*$ theory. (b) The singular
point with $\SU(p)^q\times \U(1)^{q-1}$ gauge symmetry which descends to
the $\SU(p)$ vacuum of $\N=1^*$ theory.}
\end{figure}
In Fig.~3(a), there is a $\U(1)^{N-1}$
gauge symmetry with massless hypermultiplets from strings stretching
between two four-branes on either side of the five-brane, meeting each other
at the same spacetime point. Each
such hypermultiplet carries charges under {\it two\/} different
$\U(1)$ factors. Thus in the low energy theory there are light states
charged under each of the $N-1$ $\U(1)$ factors.
Soft breaking to $\N=1^*$ will cause all these states to
condense, as is apparent from the expressions for $\Phi^\pm$ in
\eqref{ggs},
and therefore Higgses all the $\U(1)$ factors. This is the Higgs
vacuum. Fig.~3(b) illustrates a more general situation where the branes
are arranged in $p$ coincident helices, where $p$ is a divisor of
$N$. The gauge symmetry is $\SU(p)^q\times \U(1)^{q-1}$ with $pq=N$. The
massless strings stretching across the five-brane,
between two stacks of D4-branes meeting each other, are bi-fundamentals
under the
two $\SU(p)$ factors and charged under the two $\U(1)$'s associated with
the motion the two stacks. Soft breaking to $\N=1^*$
breaks the $\U(1)$ factors and simultaneously Higgses the off-diagonal
parts of the non-Abelian gauge symmetry leaving only an unbroken
$\SU(p)$ gauge group with $\N=1$ supersymmetry. Since the low energy
effective theory is $\N=1$ supersymmetric gauge theory with gauge
group $\SU(p)$, each such
configuration gives $p$ vacua, so that the total number of massive vacua of
$\N=1^*$ theory is given by the sum over divisors of $N$.

We can now extend the above picture fairly straightforwardly to the $\N=1^*$
deformation of the $\N=2$ elliptic models. As far as the massive vacua
are concerned once again we expect the relevant
Coulomb branch singularities to be those where a maximal number of
massless hypermultiplets appear, Higgsing all the classically visible
$\U(1)$ factors upon soft breaking to $\N=1$ supersymmetry.
At these positions
the branes arrange themselves in $k$ helices as shown in
Fig.~4.
\begin{figure}
\begin{center}
\mbox{\epsfxsize=2.5in\epsfbox{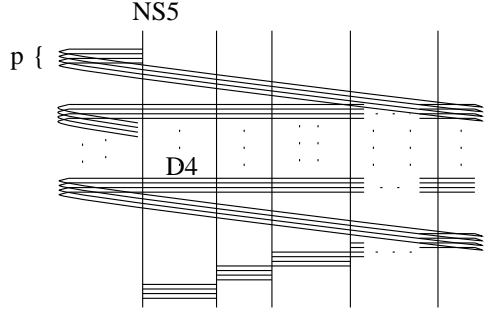}}
\end{center}
\caption{\small A Coulomb branch singularity of the elliptic model with
$\SU(p)^{kq}\times \U(1)^{k(q-1)}$ gauge symmetry. Upon soft breaking to
$\N=1$ SUSY it yields $p^k$ massive vacua.}
\end{figure}
As in the $\N=1^*$ theory, the global mass sets the pitch of a helix
as we go around the $x^6$ circle.
We can choose the shift of $m$ to be anywhere, for instance
between the $k^{\rm th}$ and $1^{\rm st}$ five-brane. The massive vacua
correspond to precisely $k$ helices one of which begins, and one ends, on each
of the $k$ five-branes. Any fewer than $k$ helices and one cannot accommodate
the $k$ hypermultiplet masses. Any more, and there would be
additional moduli in the configuration leading to a massless vacuum
upon soft breaking to $\N=1$.
In general each of the $k$ helices can be composite consisting of a number of
coincident $D4$-branes, yielding
unbroken non-Abelian factors. For a massive vacuum it is
necessary that the number in each helix be the same, namely
$p$ a divisor of $N$: $p\,q=N$. The low energy theory at such a point
(as in Fig.~4) on
the moduli space of the $\N=2^*$ quiver theory has $\SU(p)^{kq}\times
\U(1)^{k(q-1)}$ gauge symmetry with $k(q-1)$ massless
hypermultiplets. The latter originate, as before, from massless strings
ending on two coincident four-brane stacks, one on either side of an NS5.
The $\N=1^*$ deformation leads to the
condensation of these massless hypermultiplets. This not only Higgses
the classically visible $\U(1)$'s, but also breaks the product gauge
group associated with each composite helix to its
diagonal $\SU(p)$ subgroup. As there are $k$
composite helices, this leaves a classically
unbroken $\SU(p)^k$ gauge theory with massive matter and $\N=1$ supersymmetry
which yields $p^k$ massive vacua in the quantum
theory. There is an additional degeneracy that arises in the following way.
There are $q=N/p$ stacks of D4-branes intersecting
a five-brane from either side. Since there is one helix that starts and
one that ends on each five-brane, $q-1$ of the stacks are continuous
across the five-brane but one has a jump. As the jump may be chosen to be
at any of the $q$ stacks there is an additional degeneracy of $q$ per
five-brane. Actually the total degeneracy turns out to be $q^{k-1}$
since the jump at one of the five-branes can be fixed without-loss-of-generality.
Thus the total number of massive vacua of the $\N=1^*$ quiver theory is
\EQ{
\#\big(\text{massive vacua of }\N=1^*\text{ quiver}
\big)=\sum_{p|N}p^kq^{k-1}=N^{k-1}\,\sum_{p|N}p\ .
\label{massivevacua}
}
In the following sections, we will show how the exact solution of the elliptic
model can be used to reproduce this vacuum counting in the quantum
theory. In addition the picture
described above in terms of ``four-brane helices'' will have a natural
origin in the quantum solution to be described from the M-theory
point-of-view.

{\it Systematic semi-classical analysis}

After the intuitive discussion explored above,
we now consider the problem of the
vacuum structure of the quiver theories more systematically.
As in the $\N=1^*$ case, we first consider the classical vacua that
arise from the tree-level superpotential \eqref{suppottree}.
This can be re-cast as
\EQ{
W=\frac1{g^2}{\rm Tr}\Big\{(\Phi+X)[\Phi^+,\Phi^-]-
\frac m{k}\Phi^+\Phi^--\sum_i\mu_i\Phi^2_i\Big\}
\ .
}
Here, $X$ is a diagonal matrix with $k$
independent parameters $x_i$
corresponding to the VEVs of the $\U(1)$-components of the $\U(N)$
gauge groups which are frozen in the infra-red. These parameters are
determined in terms of the masses $m_i$ via
\EQ{
m_i-\frac m{k}=x_i-x_{i+1}\ .
\label{kkj}
}
Choosing $\sum_ix_i=0$ then fixes the $x_i$.
$\Phi$ and $\Phi^\pm$ are $kN\times kN$-dimensional matrices and if we
choose an ordering for the elements so that the $i^{\rm th}$ $\SU(N)$
factor of the gauge group is associated to the $i\ \text{mod}\ k^{\rm th}$ row
and columns, then
\EQ{
X={\rm
diag}(x_1,\ldots,x_k,x_1,\ldots,x_k,\ldots\ldots,x_1,\ldots,x_k)\ .
\label{ord}
}
With this ordering $\Phi$ and $\Phi^\pm$
can only have non-zero elements in positions $(u,v)$, $1\leq u,v\leq Nk$:
\EQ{
\Phi:\qquad \big(u,u+nk\big)\ ,\qquad\Phi^\pm:\qquad
\big(u,u\pm1+nk\big)\ ,\qquad n\in{\mathbb Z}\ .
\label{llls}
}
We also define $\Phi_i$, $i=1,\ldots,k$ as the elements of $\Phi$
pertaining to the $i^{\rm th}$ $\SU(N)$ gauge group factor. The matrix
$\Phi_i$ has non-zero elements in position $(i+mk,i+nk)$,
$m,n\in{\mathbb Z}$.
The classical vacuum structure of the $\N=1^*$ deformation is obtained
from the $F$-flatness conditions
\AL{
[\Phi+X,\Phi^\pm]&=\frac m{k}\Phi^\pm\ ,\label{hha}\\
\P\Big\{[\Phi^+,\Phi^-]\Big\}&=2\sum_i\mu_i\Phi_i\ ,\label{hhb}
}
modulo complex $\SL(N,{\mathbb C})^k$ gauge transformation.
In \eqref{hhb} $\P$ is a projection onto the traceless part of each of
the $\U(N)$ factors.

Complex gauge transformations can be used to
diagonalize $\Phi$. This uses up all of the complex gauge
transformations apart from the abelian transformations that fix
diagonal matrices and the Weyl group acting as permutations.
The solutions of \eqref{hha} are
associated to the partitions
of the {\it ordered\/} set of $kN$ objects which
are identified with the diagonal elements of $\Phi$:
\SP{
&
\Big\{1,2,\ldots,k,1,2,\ldots,k,\ldots\ldots,1,2\ldots,k\Big\}\\
&\qquad\qquad\qquad\qquad\qquad
\longrightarrow\Big\{\underbrace{1,\ldots,i_1}_{{\EuScript A}_1}\Bigg|
\underbrace{i_1+1,\ldots,i_2}_{{\EuScript A}_2}\Bigg|
\ldots\ldots\Bigg|\underbrace{i_{n-1}+1,\ldots,
k}_{{\EuScript A}_n}\Big\}\ .
\label{part}
}
Here the labels $1,2,\ldots,k$ refer to the $k$ $\SU(N)$ factors. The
subsets ${\EuScript A}_r$ are themselves ordered sets
and we will define
\EQ{
\ell_r={\rm dim}\,{\EuScript A}_r\ ,\qquad\sum_{r=1}^n\ell_r=Nk\ .
}
The labelling on the subsets ${\EuScript A}_r$ is irrelevant.
In other words, a partition giving subsets
${\EuScript A}'_r$ is equivalent to the first
if they are some re-labelling of the
${\EuScript A}_r$. This equivalence is due to
the action of the
Weyl group of $\SU(N)^k$. On the ordered set of $kN$ objects, the
Weyl group of the $i^{\rm th}$ $\SU(N)$ gauge group factor
acts by permuting the $i^{\rm th}$, $i+k^{\rm th}$, $\ldots$,
$i+(kN-1)^{\rm th}$ elements:
\[
\Big\{1,\ldots,\underset\uparrow i,
\ldots,k,1,\ldots,\underset\uparrow
i,\ldots,k,\ldots\ldots,1,\ldots,\underset\uparrow i,
\ldots,k\Big\}\ .
\]
For example, take $\SU(2)^2$. The partition
$\big\{\{1,2\},\{1\},\{2\}\big\}$ is equivalent to
$\big\{\{1\},\{2\},\{1,2\}\big\}$ but inequivalent from
$\big\{\{1\},\{2,1\},\{2\}\big\}$.
In addition to these equivalences,
there is a restriction on the allowed
partitions, proved below,
which requires that the set\footnote{Notice $i_n$ is fixed to be $k$.}
\EQ{
\Big\{i_1,i_2,\ldots,i_n\Big\}\supset\Big\{1,2,\ldots,k\Big\}\ .
\label{alpt}
}
So in $\SU(2)^2$, $\big\{\{1,2,1,2\}\big\}$ and
$\big\{\{1,2\},\{1,2\}\big\}$ are {\it not\/} allowed
partitions, but $\big\{\{1,2,1\},\{2\}\big\}$, for instance, is allowed.

For a given allowed partition
\EQ{
\Phi+X=\frac m{k}\MAT{c_11_{\sst[\ell_1]\times[\ell_1]}+
J_3^{(\ell_1)}&&&\\ &c_21_{\sst[\ell_2]\times[\ell_2]}+J_3^{(\ell_2)}&&\\
&&\ddots&\\
&&&c_n1_{\sst[\ell_n]\times[\ell_n]}+J_3^{(\ell_n)}}\ ,
\label{xxc}
}
where $J_3^{(\ell)}$ is the diagonal generator of the $\ell$-dimensional
representation of $\SU(2)$.
The solution can be visualized in the Type IIA brane picture as $n$
helices, the $r^{\rm th}$ having
$\ell_r/k$ complete rotations around $x^6$ and a centre-of-mass in the
$v$-plane at $c_r$.
Notice that the pitch of a helix as one
goes around $x^6$ is $m$, where
$m$ is the $\N=2^*$ global mass. The $\Phi^\pm$ share the same block
structure:
\EQ{
\Phi^\pm=\begin{pmatrix}K^\pm_1&&&\\
&K^\pm_2&& \\ &&\ddots&\\ &&&
K^\pm_n\end{pmatrix}\ ,
}
where the blocks $K^\pm_r$ have the same pattern of
non-vanishing elements
as $J_1^{(\ell_r)}\pm iJ_2^{(\ell_r)}$, the conventional raising
and lowering matrices of the $\ell_r$-dimensional
$\SU(2)$ representation. However, unlike in the $\N=1^*$ case,
the numerical values are not the same. Nevertheless,
we can use the remaining abelian symmetries to set
\EQ{
K^-_r=\begin{pmatrix}0&&&\\ 1 &0&&\\ &\ddots&\ddots&\\
&&1&0\end{pmatrix}
\label{qwa}
}
leaving
\EQ{
K^+_r=\begin{pmatrix}0&k^{(r)}_1&&\\ &\ddots&\ddots&\\
&&0&k^{(r)}_{\ell_r-1}\\
&&&0\end{pmatrix}
\label{qwb}
}
for constants $k^{(r)}_l$, $l=1,\ldots,\ell_r-1$.
Notice that the non-zero elements of $\Phi$ and
$\Phi^\pm$ are consistent with the grading \eqref{llls}.

The restriction on the allowed partitions \eqref{alpt} means that that
there must be at least one block of the partition that ends on
each of the $k$ gauge group factors, {\it i.e.\/}~at least one of the
$i_r$ equals each of $1,2,\ldots,k$.
Suppose this were not true for
the $i^{\rm th}$ gauge group factor. By taking the difference of the
trace of \eqref{xxc} for the $i^{\rm th}$ and
$i+1^{\rm th}$ gauge group factors, one would find
\EQ{
x_i-x_{i+1}=-\frac mk\ ,
}
in contradiction with $m_i$ being generic in \eqref{kkj}. In particular,
notice that this means the allowed partitions must have $n\geq k$; in other
words there must be a least $k$ helices. Notice that in terms of the
Type IIA diagrams that the condition means that at least one helix
must end (and consequently at least one must start) on each NS5-brane.

Finally, we must solve \eqref{hhb}. This equation can be re-cast as
\EQ{
[\Phi^+,\Phi^-]=2\sum_i\mu_i\Phi_i+A\ ,
\label{jju}
}
where $A$ is accounts for the trace parts of the $\U(N)$ factors:
\EQ{
A={\rm
diag}(a_1,\ldots,a_k,a_1,\ldots,a_k,\ldots\ldots,a_1,\ldots,a_k)\ ,
}
for, as yet, arbitrary constants $a_i$.
All in all, we have $n+k+kN-n$ complex unknowns in $c_r$,
$r=1,\ldots,n$, $a_i$, $i=1,\ldots,k$ and
$k^{(r)}_l$, $l=1,\ldots,\ell_r-1$ and $r=1,\ldots,n$. 
These are subject to the following
conditions. Firstly, $k$ complex conditions arise from the tracelessness of
$\Phi$, \eqref{xxc}, in each of the $\SU(N)$ factors. Secondly
Eq.~\eqref{jju} gives $Nk$ complex conditions. So there are $(N+1)k$
complex conditions for $(N+1)k$ complex unknowns and so generically
there exists a solution for each allowed partition.
Generically, the solutions have no
additional moduli and so the low energy theories will be pure $\N=1$
supersymmetric gauge theory with a gauge group that is
the subgroup of $\SU(N)^k$ fixing $\Phi$ and $\Phi^\pm$.
A given allowed partition generically leads to a solution with
unbroken $\U(1)$ factors and so to a massless, or Coulomb,
vacuum. However, for very special partitions, the unbroken gauge group
is empty or non-abelian giving Higgs and confining vacua with a mass gap.
We now identify these massive vacua. Firstly, there are
partitions into $n=k$, the smallest number of possible, subsets:
\EQ{
\text{Higgs Partition}=\Big\{\A_1,\A_2,\ldots,\A_k\Big\}\ .
}
For these vacua, the unbroken gauge
group is empty, so they are Higgs vacua. The number of them can be
determined as follows. Since $n=k$, there is a single $i_r$ in
\eqref{part} associated to each of the $k$ gauge group factors. A
given $i_r$ can therefore be situated in one of $N$ places. However,
$i_k$ is fixed, so the total degeneracy of Higgs vacua is $N^{k-1}$.
The partition with into $Nk$ sets ${\EuScript A}_r$ with a single
element each corresponds to $\Phi=\Phi^\pm=0$.
This vacuum leaves unbroken the whole
$\SU(N)^k$ gauge group, and since each $\SU(N)$ factor yields $N$
confining quantum vacua, overall there are $N^k$ massive vacua of
this type. More generally there are a number of massive vacua
associated to each divisor of $N$, $N=pq$. The associated partition
are of the form
\EQ{
\text{Confining Partition}=
\Big\{\underbrace{\A_1,\ldots,\A_1}_{p\text{ times}},
\underbrace{\A_2,\ldots,\A_2}_{p\text{ times}},\ldots\ldots,
\underbrace{\A_k,\ldots,\A_k}_{p\text{ times}}\Big\}\ ,
}
where $\sum_{i=1}^k\ell_i=qk$ and $\big\{\A_1,\ldots,\A_k\big\}$ is an
allowed---Higgs vacuum---partition of an $\SU(q)^k$ theory. The
unbroken gauge group is $\SU(p)^k$, giving a quantum degeneracy of
$p^k$, and since there are $q^{k-1}$ inequivalent Higgs vacua
for an $\SU(q)^k$ theory, the total number of massive vacua
associated to the divisor $p$ of $N$, is $q^{k-1}p^k=N^{k-1}p$. This gives the
total number of massive vacua as in Eq.~\eqref{massivevacua}. Hence, our more
intuitive discussion in terms of helices produces the right
combinatorics as this more detailed semi-classical analysis.

An example is in order. Consider the $\SU(2)^2$ theory. The vacua are
associated to the partitions of the ordered set $\big\{1,2,1,2\big\}$.
The partitions $\big\{\{1,2,1,2\}\big\}$ and
$\big\{\{1,2\},\{1,2\}\big\}$ are ruled out by the
restriction \eqref{alpt}. This leaves: (i) $\big\{\{1,2,1\},\{2\}\big\}$
and $\big\{\{1\},\{2,1,2\}\big\}$
which correspond to two inequivalent Higgs vacua; (ii)
$\big\{\{1,2\},\{1\},\{2\}\}$,
$\big\{\{1\},\{2\},\{2,1\}\big\}$ which have unbroken gauge group
$\U(1)$ and so are massless; and (iii) $\big\{\{1\},\{2\},\{1\},\{2\}\big\}$
which leads to an unbroken
$\SU(2)^2$ gauge symmetry and four massive confining vacua in the
quantum theory. Overall there are
six massive vacua which is consistent with \eqref{massivevacua} and
two additional massless vacua.

{\it Phase structure}

We end this section on the (semi-)classical analysis of the $\N=1^*$
quiver theory with some remarks on the phase structure of the massive
vacua that we found above. Each $\SU(N)$ factor in the quiver
has matter fields in the fundamental representation which therefore
transform under the center of that $\SU(N)$ factor. However the matter
fields are neutral under a
simultaneous gauge rotation in all the $\SU(N)$ factors by an element
of the center $U=e^{2\pi i a\over N}1_{\sst[N]\times[N]}$,
$a=0,\ldots,N-1$. This can
be interpreted as the
center of the diagonal gauge group $\SU(N)_D$ under which all fields
appear to transform in the adjoint representation. This may then be used to
classify the phases of the $\N=1^*$ quiver theory. However since
there is only one kind of ${\mathbb Z}_N$ symmetry---the
center $({\mathbb Z}_N)_D$ of the
diagonal $\SU(N)$, the number of possible phases is much smaller than
the number of massive vacua. As in the
case of the basic $\N=1^*$ theory \cite{donwitt,polstr}, the possible
phases are given by the order $N$ subgroups of
$({\mathbb Z}_N)_D\times({\mathbb Z}_N)_D$. The latter characterizes the
$({\mathbb Z}_N)_D$-valued electric $(n_e,0)$ and magnetic $(0,n_m)$ charges
respectively,
of the sources used to
probe the theory. This implies, for instance, that all the vacua with
unbroken $\SU(p)^k$ gauge symmetry may be classified into $p$ distinct
phases with $N^{k-1}$ vacua in each phase. These $p$ phases
correspond to the subgroups of $({\mathbb Z}_N)_D\times({\mathbb
Z}_N)_D$, generated by
the screened charges $(p,0)$ and $(l,q)$ with $pq=N$ and
$l=0,1,\ldots,p-1$. The $N^{k-1}$ Higgs vacua correspond to $p=1$.
Apart from computing the condensates, there is
however no physical way to distinguish the $N^{k-1}$ vacua in a given phase of
the diagonal gauge group. Recall that there are two sources of the
degeneracy of vacua in a given phase of the diagonal gauge group:
(i) the multiplicity $p^{k-1}$ associated with the Witten index for the
remaining
$k-1$ $\SU(p)$ factors (excluding the diagonal one); and (ii) a factor of
$q^{k-1}$ associated to the
position of the jumps in the four-brane helices. In the limit of
infinitely massive matter fields, the former may be associated with
the emergence of an accidental ${\mathbb Z}_{2p}$ symmetry in each $\SU(p)$
factor which is spontaneously broken to ${\mathbb Z}_2$ by gaugino condensation
in each factor. The additional degeneracy of $q^{k-1}$ has a more
unusual interpretation which we return to later.

\section{Lifting to M Theory}

\subsection{Review of the solution to the elliptic model}

We now consider the solution of the elliptic model, {\it i.e.\/}~the $\N=2^*$
quiver theory with the goal of extracting exact results for the
holomorphic observables and vacuum structure of the $\N=1^*$ quiver
theory. To this end, we will follow the prescription outlined in
\cite{wittm} and begin with a brief review.

The exact low energy  effective action for the elliptic model can be
obtained from the Type IIA brane setup in the limit of large string
coupling $g_s\gg1$ accompanied by a simultaneous rescaling of the
five-brane separations so that the $k$
gauge couplings $\{\tau_i\}$ are unchanged. In this limit the
compact M-dimension $x^{10}$ opens up and the system
of intersecting branes in Type IIA lifts to a single smooth
M5-brane in M-theory. There are now two circles in spacetime: the
compact $x^6$ dimension with radius $L$ and the M-theory dimension
$x^{10}$ with radius $R$. The locations of
the IIA NS5-branes are conveniently described by the holomorphic
coordinate $z=x^6+ix^{10}$. The $(v,z)$ space is locally ${\mathbb C}\times
E_{\tau}$, with $v\in{\mathbb C},\;z\in E_\tau$ and $E_{\tau}$ a genus one
Riemann surface. Specifically, the complex structure $\tau$ of the spacetime
torus $E_{\tau}$ is determined by obtaining it as a quotient of the
$z$-plane by the two equivalences
\SP{
\text{(i)}&\qquad\qquad x^6\rightarrow x^6+2\pi L, \;{\rm combined\;with}\;\;
x^{10}\rightarrow x^{10}+\theta R,\\
\text{(ii)}&\qquad\qquad x^{10}\rightarrow x^{10}+2\pi R\ .
\label{squotient}
}
Thus $E_{\tau}$ has complex structure $\tau=iL/R +\theta/2\pi$ which is
the coupling associated with the overall $\U(1)$ factor in the IIA
setup; $L/R=4\pi/g^2=4\pi\sum_i 1/g^2_i$ and $\theta=\sum_i\theta_i$.
The $x^{10}$-positions of the IIA NS5-branes can be directly related to
the theta-angles of each gauge group factor in
the quiver,
\EQ{
\theta_i={x^{10}_{i+1}-x^{10}_i\over R}\ .
\label{thetangles}
}
Thus the M-theory spacetime contains a torus $E_\tau$
with complex structure $\tau$ and $k$ distinct, unordered marked
points $z_i$, $i=1,\ldots,k$, to be identified
with the positions of the NS5-branes. Different choices of $\tau$ and
variations in the positions of the NS5-branes lead to different gauge
couplings and theta angles for the $\SU(N)$ factors in the gauge
group; explicitly
\EQ{
\tau_i=\frac{i(z_{i+1}-z_i)}{2\pi R}\ ,
}
where the $z_i$'s are labelled according to their order around $x^6$.
In the following, for convenience, 
we re-scale the torus $E_\tau$ so that the M-theory
radius $R$ is unity. This makes the $z_i$'s dimensionless.

In the presence of the overall mass shift $m$ described earlier, the
$(v,z)$ space in which the M5-brane propagates is to be thought of as a
${\mathbb C}$ bundle over $E_{\tau}$ which is defined as a quotient ${\mathbb
Q}_m$ obtained by
dividing by the combined operations
\SP{
&x^6\rightarrow x^6+2\pi L,\\
&x^{10}\rightarrow x^{10}+\theta R,\\
&v\rightarrow v+m.
\label{vsquotient}
}
This twisting of the spacetime can be undone almost everywhere at the
expense of a special point on
$E_\tau$, the location of which is arbitrary, at which certain
discontinuities will occur. Although arbitrary, this point (chosen at
$z=0$ below) will appear
in the solution of the model as we see below.

The M5-brane itself has world-volume $\Sigma\times{\mathbb R}^{1,3}$ where
$\Sigma$ is a Riemann surface embedded in ${\mathbb Q}_m$. This is the
Seiberg-Witten curve for the theory. The Type IIA setup of
intersecting branes naturally lifts to a Riemann surface $\Sigma$ with genus
$(N-1)k+1$ which is precisely the rank of the low-energy gauge group.
{\it More importantly, the curve $\Sigma$ is a branched
$N$-fold cover of the torus $E_{\tau}$}. This is clear from the Type
IIA limit wherein the curve reproduces $N$ D4-branes, each wrapped
around $E_\tau$. This curve is described by the equation
\EQ{
F(z,v)=v^N-f_1(z)v^{N-1}+f_2(z)v^{N-2}+\ldots+
(-1)^Nf_N(z)=\prod_{a=1}^N(v-v_a(z))=0\ ,
\label{curve}
}
with $z\in E_\tau\simeq{\mathbb C}/\Gamma$ where $\Gamma$ is the lattice
$2\omega_1{\mathbb Z}\oplus2\omega_2{\mathbb Z}$ with
$\omega_2/\omega_1=\tau$.\footnote{Note that our
parameterization of the curve differs from that presented in \cite{wittm}
which was written in terms of variables $x$ and $y$ specifying
a point on $E_\tau$ via the Weierstrass equation
$y^2=4x^3-g_2(\tau;\omega_1,\omega_2)x-g_3(\tau;\omega_1,\omega_2)$.
This is in fact solved by $x=\wp(z;\omega_1,\omega_2)$ and
$y=\wp^\prime(z;\omega_1,\omega_2)$ where
$\wp(z;\omega_1,\omega_2)$ is the Weierstrass function.} Thus, for every
point $z$ in $E_\tau$, the equation \eqref{curve} has $N$ roots representing
the $v$-coordinates of the $N$ D4-branes. The locations $z_i$ of simple
poles of the $v_a(z)$ represent the positions of the NS5-branes. Let
us now recall in detail following Witten \cite{wittm}
the properties of the functions
$v_a(z)$ and $f_a(z)$.

The functions $f_a(z)$ are: (i) single-valued elliptic functions on
$E_\tau$ with only simple poles at $z=z_i$ and a pole of order $a$ at
$z=0$; and (ii) the singularity at $z=0$ may be removed by a linear 
re-definition of $v$. These imply that the $v_a(z)$ which need not
be single-valued on $E_\tau$ must have the properties: (i)
that near $z=z_i$, exactly one of the $v_a(z)$ has a simple pole; and (ii)
all the $v_a$'s have a simple pole at $z=0$ (the special point of
discontinuity described above) with exactly the same residue.

The parameters of the theory, namely the hypermultiplet bare
masses $m_i$ and the order parameters on the Coulomb branch of the $\N=2$
elliptic model must be encoded in the solution $F(z,v)=0$.
The hypermultiplet bare masses are contained in the singular part of
$f_1(z)=\sum_a v_a(z)$. In particular $m_i$ is equal to $1/N$ times
the residue at $z=z_i$ (the location of an NS5-brane)
\cite{wittm,donwitt}.\footnote{In the parameterization of
\cite{wittm} the masses are $1/N$ times the residues of the differential form
$f_1(x,y)dx/y$ which is precisely
the residue of $f_1(z)dz$ in our
notation.} Since $f_1(z)$ is meromorphic
on $E_\tau$, the sum of its residues must vanish and therefore it
must have yet another simple pole which may be chosen
at $z=0$ with residue $N\sum_i m_i=Nm$ incorporating the global
mass. The constant part of $f_1(z)$ is the order parameter for the
overall $\U(1)$ factor in the elliptic model.
Each of the remaining $N-1$ functions $f_a(z)$ have $k-1$ independent residues
at
the simple poles and a constant piece leading to a total of $(N-1)k$
parameters. This is consistent with the Riemann-Roch Theorem which
determines the space of meromorphic functions on $E_\tau$ with $k$
distinct simple poles to be $k$-dimensional. These $(N-1)k$ parameters
are to be identified with the order parameters on the Coulomb branch of
the $\SU(N)^k$ elliptic model. In the following sections our goal will
be to compute these order parameters at the special singular points on
the Coulomb branch which  descend to ${\cal N}=1$ vacua upon mass
perturbation of the theory.

\subsection{Duality symmetries}

The $\N=2$ quiver theories (without hypermultiplet masses)
have a nontrivial duality symmetry that
is a generalization of the $\SL(2,{\mathbb Z})$ duality
group of the ${\cal N}=4$ theory. This duality symmetry is manifest
in the Type IIA/M-theory construction of the quiver theory at its
conformal point where all VEVs vanish. At this point the $N$ D4-branes
lift to an M5-brane which is multiply wrapped around the M-theory
spacetime torus $E_\tau$. The low-energy four dimensional field theory
then inherits the duality symmetry of M-theory under the action of
$\SL(2,{\mathbb Z})$ on the complex structure of the spacetime torus
$E_\tau$. This duality acts on the overall gauge coupling $\tau$ in
the usual way and may be interpreted as modular invariance of the
diagonal $\SU(N)$ factor.

However, in addition to this the
quiver theory has a much more unusual symmetry at the origin of the
Coulomb branch which we now describe. The $k$ individual gauge-couplings are
encoded in
the complex structure of $E_\tau$ and the positions of the $k$
indistinguishable marked points $z_i$, representing the five-brane
positions. Moving a given NS5-brane all the way around one of the
cycles of the torus (with
the remaining $k-1$ NS5's fixed), leaves the brane configuration
unaltered and therefore is a symmetry of the low energy
theory. But this operation changes the gauge couplings of the
individual $\SU(N)$ factors in the quiver as they are given by
separations of the five-branes. The positions of the latter
change according to
\SP{
&z_i\rightarrow
z_i+2m\omega_1+2n\omega_2;\;\;m,n\in{\mathbb Z}\ ,\\
&z_l \rightarrow z_l, \;\;{\rm all}\;l\neq i\ .
\label{newduality}
}
Clearly this transformation includes the familiar shift in the theta angles
$\theta_i\rightarrow\theta_i+2m\pi$ and
$\theta_{i+1}\rightarrow\theta_{i+1}+2m\pi$ which is a symmetry of the
theory. However, in addition the transformation also implies a
periodicity of the inverse gauge couplings $1/g^2_i$ with period $1/g^2$. 
In the T-dual Type IIB picture of the quiver theory, the periodicity of the
inverse gauge couplings arises from the periodicity of the scalar fields
obtained by integrating the NS-NS 2-form over the vanishing 2-cycles at the
orbifold singularity \cite{polchinski,katz}.

As in the case of the $\N=4$ theory the duality group must be
``spontaneously broken'' at a generic point on the Coulomb branch of the
$A_{k-1}$ quiver theory. Specifically, the Coulomb branch of the quiver theory
contains certain special singular points. These are points which
become $\N=1$ vacua upon perturbing the theory with non-zero
hypermultiplet and adjoint masses, $m_i\neq 0$ and $\mu_i\neq 0$. At these
singularities new massless BPS states appear in the low energy theory
and this is signalled by the degeneration of the corresponding cycles of the
Seiberg-Witten curve. As in the case of
mass deformed $\N=4$ SYM, one expects each massive vacuum of
the $\N=1^*$ quiver theory to be
associated with the condensation of a specific BPS state belonging to
the spectrum of the parent conformal quiver theory.
As the generalized duality symmetry of the quiver model described
above must leave the spectrum of the conformal theory invariant,
it therefore acts on the BPS states in the spectrum via permutation.
This in turn implies that the action of the duality group takes the
$\N=1^*$ quiver theory from one vacuum to another since each of these
vacua is associated with the condensation of a specific BPS particle in
the theory.

In addition to the above duality properties, we will argue below that
the low-energy physics, in particular the chiral sector of the
$\N=1^*$ quiver theory in {\it each vacuum} is invariant under a
symmetry which we will refer to as extended $\tilde S$-duality.

\subsection{The quantum vacuum structure of the $\N=1^*$ quiver theory}

We now turn to the description of the quantum vacuum structure
of the $\N=1^*$ quiver theory. As mentioned above, and described in
Appendix B from the point of view of brane rotation, the massive vacua of the
theory correspond to the singular points on the Coulomb branch of the
$\N=2^*$ quiver theory where the Seiberg-Witten curve undergoes
maximal degeneration to a surface of genus one.

The locations of these special singular points on the $\N=2$ moduli
space are completely specified by the $k(N-1)$ gauge-invariant order
parameters $u^{(i)}_a\equiv\langle\Tr\,\Phi^a_i\rangle$, $i=1,2,\ldots k$, and
$a=2,\ldots,N$.\footnote{Here, we use the same symbol to denote the
chiral superfield and its lowest component.}
As we have seen above, these order
parameters are encoded in the curves for the theory (modulo
ambiguities to be discussed later). The values of these order
parameters at the singular points
(as a function of the $k$ couplings $\tau_i$
and mass parameters $m_i$) yield the condensates of the $\N=1^*$
quiver theory and provide complete information on the chiral sector
of that theory.

At the special singular points which descend to
massive $\N=1$ vacua, the Seiberg-Witten curve degenerates maximally to an
unbranched (unramified) $N$-fold cover of the spacetime torus
$E_\tau$.\footnote{Note
usually in $\N=2$ theories as in \cite{swone} the curves
maximally degenerate into a sphere. In the elliptic models the curves
degenerate at most into a genus one Riemann surface. In the M-theory
construction this reflects the presence of the unbroken and decoupled
$\U(1)$ factor.} This fact and the associated ellipticity properties
will allow us to explicitly construct the functions $v_a(z)$ at the
singular points in the moduli space of the curve Eq. \eqref{curve} and
extract the condensates for the $\N=1$ theory.

An unbranched $N$-fold cover of $E_\tau$ is also a torus, but with a
different complex structure $\ttau$. As noted in
\cite{donwitt} and \cite{oferandus} such covers are classified by 3 integers
$p$, $q$ and $l$ where $pq=N$ and $l=0,1,\ldots p-1$. They are genus one
Riemann surfaces $E_{\tilde{\tau}}$ with the complex structure
${\tilde{\tau}}=(q\tau+l)/p$. $E_{\tilde{\tau}}$ can be defined as
usual as the quotient of the complex plane by the
$\tilde{\tau}$-lattice, namely
$E_{\tilde{\tau}}\simeq{\mathbb C}/\Gamma_\ttau$. In what follows, we
focus primarily on the cases with $l=0$, since the more general
covering can be obtained by suitable modular transformations. 
Fig.~5 illustrates the case
$\Gamma_\ttau\simeq\tilde{\omega}_1{\mathbb Z}
\oplus\tilde{\omega}_2{\mathbb Z}$ with
$\tilde{\omega}_1=p\omega_1$ and $\tilde{\omega}_2=q\omega_2$ and
$\ttau=\tilde{\omega}_2/\tilde{\omega}_1$.
\begin{figure}
\begin{center}
\mbox{\epsfxsize=3in\epsfbox{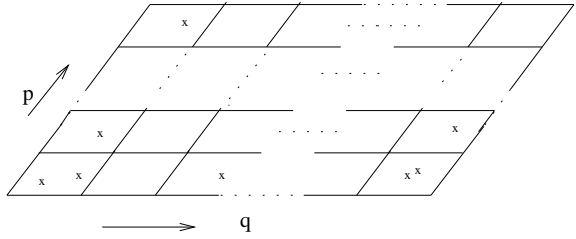}}\end{center}
\caption{\small The $N$-fold cover of $E_\tau$  with
$\ttau=q\tau/p$. The $\times$'s
represent the positions of the $k$ NS5-branes.}
\end{figure}
The fundamental parallelogram of the $\ttau$ lattice shown in Fig.~5
consists of
$N$-copies of the fundamental parallelogram of the
$\tau$-lattice with $l=0$. The NS5-branes which extend in the $012345$
directions are represented by marked points (the $\times$'s) on the
$\ttau$-torus. Each $\tau$-parallelogram represents a single Type IIA
four-brane that has grown an extra dimension in the $\omega_1$
direction (which is roughly speaking, the M-dimension).

The Seiberg-Witten curve controls the physics of
the holomorphic sector of the theory and as discussed above, at the
maximally singular points, the curve degenerates into a torus with
complex structure $\ttau$. As usual, the low-energy physics
depends only on the complex structure $\ttau$ and must be invariant
under modular transformations acting on $\ttau$. Thus VEVs of chiral
operators in each vacuum of the $\N=1^*$ quiver theory are expected
to be modular functions
of $\ttau$. This duality, referred to as $\tilde S$-duality
\cite{oferandus} in the $\N=1^*$ theory is further extended in the
case of the mass-deformed quiver theories. In addition to modular
transformations on $\ttau$, we may also move each NS5-brane
independently around non-trivial cycles of the $\ttau$-torus, inducing
corresponding shifts in the individual gauge-couplings. Such
$\ttau$-elliptic shifts in the positions of each NS5-brane are also
symmetries of each vacuum of the $\N=1^*$ quiver theory and form an
extended version of $\tilde S$-duality.

It is
worth noting that there is a simple and direct mapping
between the $\ttau$-lattice sketched in Fig. 5 and the Type
IIA picture for a massive $\SU(p)^k$ vacuum. Recall that $v_a(z)$ is
simply the $v$-coordinate of the $a$-th Type IIA four-brane expressed as a
function of
the $(x^6,x^{10})$ coordinates of the compact dimensions.
In the classical (Type IIA) limit the $x^{10}$ dimension is
vanishingly small and the $\ttau$-parallelogram describes $p$
coincident four-brane helices. In the classical limit all the NS5's are
lined along the $x^6$ direction leading to discontinuities in the four-brane
helices. Quantum effects (or the growth of the M-dimension) not only
cause the four-branes to grow an extra dimension but also lead to a
separation between the helices in the $x^{10}$-direction.

{\it Vacuum counting in the quantum theory}

Let us now see how the quantum theory reproduces the degeneracy factor of $p
N^{k-1}$ associated with each classical $\SU(p)^k$ vacuum.
Recall that the duality group of the $A_{k-1}$ quiver gauge theory
acts on the massive vacua of the $\N=1^*$ quiver model via
permutations. Thus one may sweep out all the vacua of the theory by
performing the duality operations on a given vacuum configuration in the
IIA/M-theory setup.

Firstly, for a given $p$ that divides $N$ there are $p$
inequivalent $N$-fold covers of $E_\tau$ related to one another by
$\SL(2,{\mathbb Z})$ operations on $\tau$ that sweep out the $p$ values
of $\ttau=(q\tau+l)/p$ with $l=0,1,\ldots,p-1$.
Secondly, each NS5-brane may be shifted by integer
multiples of $\omega_1$ and $\omega_2$ and placed in any
one of the $N$ $\tau$-parallelograms that make up the $N$-fold
cover. While the first operation constitutes the action of
$\SL(2,{\mathbb Z})$ on $E_\tau$, the second represents the motion of
an NS5-brane around the cycles of the spacetime torus $E_\tau$. These
operations are precisely the duality symmetries of
the parent conformal theory discussed in the previous section. Thus
they act on the vacua of the $\N=1^*$ quiver by permutation. Since
only the relative separations (and not the actual positions)
of the NS5's are physically meaningful, and each five-brane may intersect
any one of the $N$ four-branes, there are $N^{k-1}$ such distinct
configurations. Thus the total number of vacua from configurations
with a classically unbroken $\SU(p)^k$ gauge symmetry is $p N^{k-1}$. We have
already argued that the degeneracy factor of $p$ arises from $\SL(2,{\mathbb
Z})$
transformations on the coupling of the diagonal part of the gauge group.
Displacing
the NS5-branes by integer multiples of $\omega_1$ leads to a shift in
the theta angles $\theta_i$ of each gauge group factor in the
quiver and yields $p^{k-1}$ distinct configurations. Moving the
five-branes along the other period gives rise to
$q^{k-1}$ ground states. The former is in
line with our semiclassical arguments wherein a multiplicity of $p^k$
appears as a consequence of the Witten index for $\N=1$ SUSY gauge
theories with massive matter and $\SU(p)^k$ gauge symmetry. The classically
obscure degeneracy factor of
$q^{k-1}$ associated with the position of the discontinuities in the
four-brane helices in the Type IIA limit, can now be clearly understood
to be a manifestation of the non-trivial duality symmetry of the
underlying conformal quiver theory.

{\it The singular curves and condensates}

The calculation of the condensates of chiral operators in the vacua with
$\N=1$ SUSY relies on the key property that at the associated singular
points in the $\N=2$ moduli space the curve $\Sigma$ degenerates into
the torus $E_\ttau$.
This translates directly into the requirement that the
functions $v_a(z)$ describing the individual four-brane positions be invariant
under a $\ttau$-elliptic transformation {\it i.e.} under translations of $z$ by
periods of $E_\ttau$. At this point the functions
$v_a(z)$ must be thought of as
different branches of a complex function that is single-valued on
$E_\ttau$ but not on $E_\tau$. Translations by periods of the
$\tau$-torus will smoothly take us from one branch to the next
{\it i.e.} from one four-brane to the next. In particular the complete {\it set} of
functions ${v_a(z)}$ will still be invariant under periodic shifts on
$E_\tau$, but the individual $v_a$'s will get permuted by this action.
We can readily write down the explicit form of these functions at the
singular points. It will be sufficient to
consider a generic vacuum with $N=pq$ and $l=0$. All other massive
vacua may be
obtained by the operations generating the duality group of the
theory.

In a vacuum with a classically unbroken  $\SU(p)^k$ gauge symmetry, it
is convenient to denote the positions of the four-branes with two
subscripts $v_a(z)\equiv v_{sr}(z)$, $s=0,1,2,\ldots p-1;\;r=0,1,2,\ldots
q-1$. It is well-known that every elliptic function has an expansion
in terms of the Weierstrass Zeta function $\zeta(z)$ (see
Appendix A for details)
and its derivatives \cite{WW}. This function has only a simple pole at
$z=0$ with unit residue. Hence, knowledge of the singularities of an
elliptic function and associated
residues completely determines its expansion
in terms of the zeta function and its derivatives.
In addition, the Weierstrass Zeta function
$\zeta(z|\tau)$ on a torus with complex structure $\tau$ has
the following anomalous
transformation property:
\SP{\zeta(z+2\omega_\ell|\tau)=\zeta(z|\tau)+2\zeta(\omega_\ell|\tau)\ ,\qquad\ell=1,2.
}
We have argued that the four-brane positions $v_{sr}(z)$ must be
elliptic on the $\ttau$-torus. Thus they must have an expansion in
terms of the zeta function on the $\ttau$-torus, which we denote as
$\tilde{\zeta}(z)$:
\EQ{\tilde{\zeta}(z)\equiv\zeta(z|\ttau);\;\;\;
\tilde{\zeta}(z+2p\omega_1)=\tilde{\zeta}(z)+2\tilde{\zeta}(p\omega_1);\;\;\;
\tilde{\zeta}(z+2q\omega_2)=\tilde{\zeta}(z)+2\tilde{\zeta}(q\omega_2)\
.}

Based on the properties of $v_{sr}(z)$
reviewed in the Section 4.1, we also know that these functions
can only have simple poles located at the positions of the
NS5-branes with residues given by the hypermultiplet masses
$Nm_i$. Finally, we must also have that at each NS5 location, one and
only one $v_{sr}(z)$ has a simple pole. Following these
requirements, we can deduce the form of the four-brane coordinates
$v_{sr}(z)$. In particular, taking the vacuum with $l=0$
where the $i$-th
NS5-brane lies on the $(s_i,r_i)$ sheet of the covering, we have
\SP{
v_{sr}(z)=&N\sum_{i=1}^k
m_i\tilde{\zeta}(z-z_i+2(s-s_i)\omega_1+2(r-r_i)
\omega_2)\\
&-m
\Big\{\sum_{t=0}^{p-1}\sum_{u=0}^{q-1}\tilde{\zeta}(z+2t\omega_1+2u\omega_2)
+2qs\tilde{\zeta}(p\omega_1)+2pr\tilde{\zeta}(q\omega_2)\Big\}\\
&s=0,1,2,\ldots p-1;\;\; r=0,1,2,\ldots q-1.
\label{vas}
}
Note in particular that this satisfies the
requirement that at each five-brane
position, exactly one $v_{sr}(z)$, namely $v_{s_ir_i}(z)$,
has a simple pole and the residue is
given by the hypermultiplet mass: $Nm_i$. All the $v_{sr}(z)$ also have a
simple pole at $z=0$ with exactly the same residue $-m$.
Our expressions Eq.\eqref{vas} for the four-brane coordinates
$v_{sr}(z)$ are uniquely determined up to additive constants, by the
required singularity structure and the property of
$\ttau$-ellipticity. However, these additive constants do not affect
the computation of condensates of gauge-invariant chiral operators.
Due to the anomalous
transformation properties of the zeta functions, the functions
$v_{sr}(z)$ transform into one another upon going around one of the
cycles of $E_\tau$. For example, $v_{sr}(z+2\omega_1)=v_{s+1\;r}(z)$
and $v_{sr}(z+2\omega_2)=v_{s\;r+1}(z)$ where the subscripts are
defined modulo $(p,q)$ as usual. Going around the periodic dimensions takes us
smoothly from one four-brane to the next. This reflects the fact that
the Type IIA four-branes lifted to M-theory form an unbranched
(unramified) $N$-fold
cover of the torus $E_\tau$.
The other vacuum configurations with $l\neq0$,
may be obtained by using suitable
$\SL(2,{\mathbb Z})$ modular transformations on the overall coupling $\tau$.

We will now argue that the condensate $u_2^{(i)}=
\langle \Tr\,\Phi_i^2\rangle$, namely
the expectation
value of the adjoint scalar in the $i$-th $\SU(N)$ factor is encoded in
the function
\EQ{u_2(z)={1\over 2N}\sum_{s,s'=1}^p \sum_{r,r'=1}^q(v_{sr}(z)
-v_{s'r'}(z))^2.\label{phisq}}
The simplest way to see that this is  the relevant object is to first recall
the classical (Type IIA) limit. In this limit the $v_{sr}(z)$
corresponding to the four-brane positions become independent of $z$
at any given point between two adjacent five-branes. The only position
dependence in the four-brane coordinates arises from discontinuities or
breaks as we move across an NS5-brane. In the classical limit, the
positions of the D4-branes in a given gauge group factor correspond
precisely to the eigenvalues of the adjoint scalar $\Phi_i$ in that
gauge group factor. It is then straightforward to show that classically
$\langle\Tr\,\Phi_i^2\rangle={{1\over 2N}\sum_{s,s'=1}^p
\sum_{r,r'=1}^q(v_{sr}-v_{s'r'})^2}$
provided $\Phi_i$ is defined to be traceless.

When the same quantity
is lifted to M-theory it naturally acquires non-trivial
position-dependence ($z$-dependence) in that the classical
discontinuities at the five-brane
locations get smoothed out and the simple geometrical relationship
between the classical condensates $u_2^{(i)}$ and Eq \eqref{phisq} is
lost. Now $u_2$ is a smooth function of $z$ with $k$ double poles and
$k$ simple poles.
The strengths of the double poles are simply determined by
the mass parameters $m_i$, while the values of the condensates $u_2^{(i)}$ are
encoded in the residues at the simple poles and the constant part of
$u_2(z)$. The simple pole residues and the constant part of
$u_2(z)$ are moduli of the Seiberg-Witten curve and are related to the
Coulomb branch moduli of the $\N=2^*$ quiver theory.

It is easy to see that the function $u_2(z)$ is an elliptic
function on the $\tau$-torus with modular weight 2. This follows from the fact
that the ``four-brane
positions''
$v_{sr}(z)$ essentially transform into one another under $z\rightarrow
z+2m\omega_1+2n\omega_2;\;m,n\in{\mathbb Z}$, and that the $\zeta$-functions
have modular weight 1. The $\tau$-ellipticity of
$u_2(z)$ provides a natural way to separate out the singular and
constant pieces since every elliptic function has an expansion
in terms of $\zeta$-functions (see Appendix A)
\SP{u_2(z)=N(N-1)\sum_{i=1}^k
m_i^2\wp(z-z_i)+\sum_{i=1}^k\zeta(z-z_i)H_i+H_0.
\label{u2exp}}
There are only  $k-1$ independent simple pole residues $H_i$ since
we must have $\sum_{i=1}^kH_i=0$. Thus, including the constant part of $u_2(z)$
there are $k$ independent parameters which coincides with the number
of condensates $u_2^{(i)}=\langle\Tr\,\Phi_i^2\rangle$.
The connection between $u_2(z)$
and the condensates is also clearly seen via the intimate relationship
between the Coulomb branch physics of the $\N=2^*$ quiver theory and
integrable models. This aspect will be explored in great detail
in Section 4 and via a completely different physical picture, will
lead to an extremely non-trivial check of the results for the
condensates presented in this section.

At this stage it must be pointed
out that the exact relationship between the parameters $H_0$ and $H_i$
on the one hand and the physical condensates $u_2^{(i)}$ on the other, is
far from obvious. There is no unambiguous way to map the residues in
the above expressions to the VEVs of the operators $
\Tr\,\Phi_i^2$ in each gauge
group factor. In general, one expects the physical $u_2^{(i)}$ to be some
linear combination of the residues $H_i$ and $H_0$ and the identity operator.
Specifically, we
may identify a set of $k$ condensates that are modular covariant with the
correct
modular weight, respecting the symmetries of a given vacuum. However,
there is no unambiguous way to determine the exact relation between
such a set of condensates and the
VEVs of the operators $\Tr\,\Phi_i^2$.
This is a generalization of the ambiguity
encountered in the basic $\N=1^*$ case
\cite{oferandus},\cite{mismatch},\cite{nickprem}.
However although it appears difficult to obtain the condensates within a given
gauge group factor, there is a natural unambiguous combination of the
condensates
$H_0$ and $H_i$ that can be identified with the average
condensate $\tfrac1k\sum_{i=1}^ku_2^{(i)}=\tfrac1k
\sum_{i=1}^k\langle\Tr\,\Phi_i^2\rangle$. As we will argue below, this
combination will have the correct transformation properties under the
duality group and only suffer from a mild vacuum-independent additive
ambiguity {\it i.e.\/}, mixing with the identity operator.

Using Eqs. \eqref{vas} and \eqref{phisq} we may readily calculate the
residues of the function $u_2(z)$ at the simple poles $z=z_i$,
corresponding to the location of the NS5-branes.
The calculation of the $z$-independent part $H_0$ is tedious and can
be performed in two different ways. One method is to use
$\ttau$-ellipticity of
each term $(v_{sr}-v_{s'r'})^2$ in the summation in
Eq.\eqref{phisq}, expand in terms of $\tilde{\zeta}$-functions and
subsequently obtain the constant pieces by evaluating the
expressions at certain special points. The other technique is to make
use of various identities for elliptic functions provided in Appendix
A, particularly Eq.\eqref{imp}.
Below, we simply present our final results for
the condensates $H_0$ and $H_i$ for the vacuum with $l=0$ and
$(s_i,r_i)=(0,0)$. (As we have emphasized the values of the
condensates in the other massive vacua can then be deduced by moving
NS5-branes by periods of the torus and also by modular transformations
in $\tau$.)

It is convenient to introduce the following shorthand notation
for certain combinations of variables that appear in the formulas
for the condensates:
\EQ{z_{ij}\equiv
z_i-z_j\ ,\qquad\Omega_{sr}=2s\omega_1+2r\omega_2\ ,\qquad
\tilde{\wp}(z)=\wp(z|\tilde\tau)\
,\qquad\tilde{\zeta}(z)=\zeta(z|\tilde\tau)\ ,}
We then find that the simple pole residues are given by
\EQ{H_i\Big\vert_{l=0;s_i=r_i=0}=2 Nm_i\;\sum_{j\neq
i}m_j
\Big\{(N-1)\tilde\zeta(z_{ij})-\sum_{(r,s)\atop\neq(0,0)}
\big(\tilde\zeta(z_{ij}+\Omega_{sr})-\tilde\zeta(\Omega_{sr})\big)
\Big\}\ .
\label{res}}
Note that $\sum H_i=0$ as required. Note also that they are
modular functions with weight 1 under the action of $\SL(2,\mathbb Z)$
on $\tau$. Furthermore, shifting one of the NS5-branes by a period of
the $\ttau$-torus leaves the $H_i$ invariant. As argued
earlier, shifting the five-branes by periods of the $\tau$-torus has
the effect of permuting the vacua of the $\N=1^*$ quiver theory. But
since the degenerate curve at a given vacuum is simply the
$\ttau$-torus, taking an NS5-brane around one of the cycles of this
torus must be a symmetry of that vacuum and must leave the order
parameters unchanged.
Hence the $H_i$ have
the requisite properties to  be identified as condensates
characterizing a given massive vacuum of the $\N=1^*$ quiver
theory. However, it is clear from Eq.\eqref{u2exp}, that $H_0$ cannot
share this property. Moving a five-brane around the $\ttau$-torus will
necessarily lead to shifts in our definition of $H_0$ simply because
of the anomalous transformation properties of the $\zeta$-functions
characterizing the simple pole terms. This can be explicitly checked
from our expression for $H_0$:
\SP{&H_0\Big\vert_{l=0;s_i=r_i=0}=-\frac{N^2}{12}
\Big(\sum_{i=1}^km_i^2+\tfrac12\sum_{i\neq j}m_im_j\Big)
\left[E_2(\tau)-\tfrac qpE_2({\ttau})\right]\\
&+N^2\sum_{i\neq
j}m_im_j\Bigg\{\tfrac12
\big(\wp(z_{ij})-N\tilde{\wp}(z_{ij})\big)
+\frac{N-1}{2}\tilde{\zeta}^2(z_{ij})\\
&+\frac1{2N}
\sum_{(r,s)\atop\neq(r^\prime,s^\prime)}
\Big(-\tilde{\zeta}^2(z_{ij}+\Omega_{sr}-\Omega_{s'r'})+
\Big[\tfrac{2(s-s')}p\tilde\zeta(p\omega_1)+
\tfrac{2(r-r')}q\tilde\zeta(q\omega_2)\Big]^2\Big)
\\
&-\frac1N{{q\tilde{\zeta}(p\omega_1)-\zeta(\omega_1)}\over \omega_1}
z_{ij}\Big((N-1)\tilde{\zeta}(z_{ij})
-\sum_{(r,s)\atop\neq(0,0)}
\left(\tilde{\zeta}(z_{ij}+\Omega_{sr})-
\tilde{\zeta}(\Omega_{sr})\right)\Big)
\Bigg\}\ .\label{constant}}

In writing the above expression we have used the identities
Eq.~\eqref{convert}  and Eq.~\eqref{e2p}. 
Eqs.~\eqref{res} and 
\eqref{constant} are our results for $H_i$ and $H_0$ respectively,
in the vacuum with $\ttau=q\tau/p$ and with all NS5's intersecting
the same IIA four-brane. The expressions for the vacua with $l\neq 0$
may be obtained by the replacement
$\omega_2\rightarrow\omega_2+l\omega_1/q$ in the above
equations. All other massive vacua can be obtained
by displacing the five-branes by periods of the $\tau$-torus to yield
inequivalent brane configurations. As discussed earlier, for every
choice of $p,\;q$ and $l$, there are $N^{k-1}$ such vacua.

Under the duality transformation $z_1\rightarrow z_1+2p\omega_1$ the quantity
$H_0$ is not invariant, rather
$H_0\rightarrow H_0+2p\zeta(\omega_1)H_1$. This implies that $H_0$
does not respect $\tilde S$-duality and by
itself cannot be identified with the order parameter in a given vacuum.
However, there
is a linear combination of the quantities $H_0$ and $H_i$
that respects the duality symmetries of the $\N=1^*$ vacuum; namely
\EQ{H^*=H_0-{1\over k}\sum_{i\neq l}\zeta(z_{il})H_i.\label{defhstar}}
The redefined constant piece $H^*$ appears to have the properties
required of the average condensate
${1\over k}\sum_i\langle\Tr\,\Phi_i^2\rangle$.
It has dimension 2, modular
weight 2, and is invariant under permutations of the gauge group
factors and importantly is invariant under $\ttau$-elliptic shifts of
each five-brane, {\it i.e\/} $\tilde S$-duality,
which was argued to be a symmetry of each massive vacuum. As
we will see below, up to an additive vacuum-independent ambiguity, in
the classical limit it indeed corresponds to the average condensate 
${1\over k}\sum_i\langle\Tr\,\Phi_i^2\rangle$.

{\it A Superpotential for the $\N=1^*$ quiver theory}

For generic $\N=1^*$
deformations of the quiver theory, the superpotential in each vacuum of the low
energy theory takes the value
\EQ{W=-{1\over g^2}\sum_{i=1}^k\mu_i\langle\Tr\,
\Phi_i^2\rangle\ .\label{gensup}}
Our inability to map the coordinates $H_0$ and $H_i$ to the
physical condensates in each gauge group factor, implies that we
cannot determine this superpotential for generic
deformations.

However, for a subclass of deformations of the quiver theory
where all the adjoint masses are equal $\mu_i=\mu\neq 0$, the
effective superpotential in each vacuum is simply given by the
expectation value of the average condensate:
\EQ{W=-{1\over g^2}\mu\sum_{i=1}^k\langle\Tr\,\Phi_i^2\rangle=-{1\over
g^2}\mu\; k\; H^*. \label{specialsup}}
This result for the superpotential in the massive vacua of the
$\N=1^*$ quiver theory along with Eqs.\eqref{constant} and
\eqref{defhstar} are among the
central results of this paper. When $k=1$, Eq.\eqref{constant} reduces
to the expression for the
elliptic superpotential for $\N=1^*$ gauge theory obtained in
\cite{Nick,oferandus}.

A few remarks about the condensates $H^*$, $H_i$ and the
superpotential in Eq.\eqref{specialsup} are in order. Under
$\SL(2,{\mathbb Z})$ 
transformations on $\tau$, these order parameters in a vacuum with
given $p,q$ and $l$ transform with a definite modular weight into the
corresponding 
condensates in a vacuum with a different value of $p,q$ and $l$ and
with different values of the gauge couplings. For
example under $S$-duality, $H^*(p,q,l=0)$ transforms into 
$\tau^2H^*(q,p,l=0)$ and with $z_{ij}\rightarrow \tau z_{ij}$. It is
also worth noting that in the classical theory (Type IIA picture),
the
relative positions of adjacent fivebranes appear to play a special
role in that they correspond to gauge couplings of the individual
$\SU(N)$  factors in the quiver. However, the exact expressions for the
condensates in a given vacuum exhibit complete democracy in the
relative positions of any two fivebranes in the setup. 

Finally, the condensates contain a wealth of information on instanton and
``fractional instanton''  contributions in these theories. This can be
clearly seen by studying the expressions in a semiclassical expansion
{\it i.e.\/} in the $g^2,g^2_i\ll1$ limit. For example, in
this limit, the condensates in the confining vacuum with $p=N, q=1$
and $l=0$ may be expanded 
in powers of $e^{2\pi i\tau/N}$ and $e^{-z_{ij}/N}$. There is
no obvious semiclassical origin for such terms in the 4D theory as
they appear to be contributions from objects with fractional
topological charge in both the diagonal and the individual gauge group
factors. However, as seen in \cite{Nick}, in the theory on 
${\mathbb R}^3 \times S^1$, such terms can arise from semi-classical
configurations corresponding to 3D monopoles carrying fractional
topological charge.

{\it The classical limit}

We now demonstrate that $H^*$ indeed reduces to the average condensate
in the classical theory in the appropriate limit.
The classical limit of the theory is obtained by taking
$\tau\rightarrow i\infty$ which corresponds to blowing up the radius of
the $x^6$-circle, simultaneously scaling the five-brane positions so
that $\tau_i\rightarrow i\infty$ in each gauge group factor.
In this
limit we should expect our general expression for $u_2(z)$ to yield the
condensate in the $i$-th $\SU(N)$ factor provided $z_i< z <z_{i+1}$,
$z-z_i\rightarrow \infty$ and $z_{i+1}-z\rightarrow \infty$. For
simplicity, we have taken all the $\{z_i\}$ to be real.
The $\zeta$-functions then reduce to (see Appendix A):
\EQ{\zeta(z-z_j)\rightarrow -{(z-z_j)\over 12}+{1\over2}\text{sign}(z-z_j).}
while the Weierstrass function simply tends to a constant
$\wp(z-z_j)\rightarrow {1\over 12}$. Note the presence of discontinuities at the
positions of the NS5-branes in the classical limit.
The condensate in the $i$-th
$\SU(N)$ factor in the classical limit is then
\EQ{u_2^{(i)}={N(N-1)\over12}\sum_{j=1}^k m_j^2+
\sum_{j=1}^k{z_j\over 12}H_j+\sum_{j=1}^i H_j +H_0,}
so that the average condensate is
\EQ{{1\over k}\sum_{i=1}^ku_2^{(i)}={N(N-1)\over12
}\sum_{j=1}^k m_j^2+\sum_{j=1}^k{z_j\over 12}H_j
+{1\over k}\sum_{i=1}^k\sum_{j=1}^i H_j +H_0.}
It is straightforward to see that in the classical limit our
definition for $H^*$ in
Eq.\eqref{defhstar} approaches
precisely this value up to a vacuum-independent additive constant:
\EQ{H^*\rightarrow \sum_{j=1}^k{z_j\over 12}H_j
+{1\over k}\sum_{i=1}^k\sum_{j=1}^i H_j +H_0.}

The discrepancy is vacuum-independent because we can trace its origin
to the double pole terms in $u_2(z)$ involving the Weierstrass
functions. The order parameters of the theory and therefore the
vacuum-dependent information is completely encoded
in the simple poles and constant pieces, $H_i$ and $H_0$ . The
quantity $H^*$
contains precisely the right combination of these parameters that
reproduces not only the correct semiclassical limit of the average
condensate but also has the right transformation properties under the
duality symmetries.

\section{Compactification to Three Dimensions}

In the previous section, we ``solved'' for the vacuum structure of the
mass deformed theory
by lifting the Type IIA brane configuration to M Theory. As we saw,
the massive vacua have a very simple interpretation as unbranched
multiple covers of the basic underlying torus $E_\tau$. There is another way to
``solve'' the theory involving compactification on a circle to
three dimensions \cite{Nick},\cite{sw3d}. On compactification,
the $2n=2(kN-k+1)$ real
dimensional Coulomb branch of the quiver theory
is enlarged. This is
because the $\U(1)^n$ gauge field can have a non-trivial Wilson line around the
compact dimension, which we choose as $x^3$. In addition, the
three-dimensional
gauge field transforming in the unbroken $\U(1)^n$, on the
Coulomb branch, can be exchanged, under $S$-duality, for $n$ real
scalar fields. In four-dimensions, the effective couplings of the
$\U(1)^n$ theory are encoded in the $n\times n$-dimensional matrix
$\tau_{uv}$ which is precisely the period matrix of the Seiberg-Witten
Riemann surface $\Sigma$.
The Wilson line and dual photons naturally pair up
to form $n$ complex scalars $X_u$ which are periodic variables
\EQ{
X_u\thicksim X_u+2\pi im_u+2\pi i\tau_{uv}n_v\ ,\qquad
m_u,n_u\in{\mathbb Z}\ .
\label{xper}
}
This means that the Wilson line and dual photon are valued in the
Jacobian variety ${\EuScript J}(\Sigma)$. So the dimension of the Coulomb
branch, denoted $\ms$,
of the three-dimensional theory is $4n$ and can be thought of
as a fibration of the Jacobian variety ${\EuScript J}(\Sigma)$ over the
four-dimensional Coulomb branch $\Sigma$. It turns out
that $\ms$ is rather special: it is hyper-K\"ahler
space.

The low-energy effective three-dimensional theory is described by a
supersymmetric sigma model with $\ms$ as target.
When an $\N=1^*$ mass deformation is added to the four-dimensional
theory, the sigma model is perturbed by a superpotential. The phase
structure can then be determined from the superpotential. The
key point that makes the compactification approach attractive even
when one is interested in the four-dimensional theory
is that the superpotential is completely independent
of the compactification radius. Therefore the vacuum structure and the values
of condensates apply also in the de-compactification limit. By
``solving'' the model, we mean finding the exact form for the
superpotential.

For the $\N=1^*$ deformation of the $\N=4$ theory, the form of the
superpotential was arrived at from purely physical reasoning: by
imposing all the symmetries of the low-energy three-dimensional
theory \cite{Nick}. In addition, in the limit of small radius, semi-classical
reasoning applies and provides a strong constraint on the functional
form of the superpotential. However, there is a more direct way to
construct the superpotential which relies on the fact that the
Seiberg-Witten fibration, and hence $\ms$, is the phase space of a
complexification of a classically
integrable dynamical system\footnote{This means that the dynamical
variables, Hamiltonians, {\it etc.\/}, are taken to be complex.} (for
a recent thorough review of this connection
see \cite{Gorsky:2000px} and references therein). In
this description,
the Coulomb branch of the four-dimensional theory
is parameterized by the set of
$n$ (complex) Hamiltonians and the fibre is parameterized by
$n$ (complex) angle variables. The $\N=1^*$ deforming operator
is then associated with a particular
combination of the Hamiltonians and this gives the
superpotential. The problem of finding the explicit superpotential is
then the problem of finding a convenient and unconstrained
parameterization of the phase
space $\ms$ and then finding the appropriate Hamiltonian describing
the $\N=1^*$ deformation.

\subsection{The three-dimensional Coulomb branch}

The first problem is to find a convenient description of the hyper-K\"ahler
manifold $\ms$. Remarkably, following the approach adopted by Kapustin
\cite{Kapustin:1998xn}, this can be achieved by using a version of mirror
symmetry. What results is a description of $\ms$ in terms of a
hyper-K\"ahler quotient \cite{HKQ} which can be made very explicit.

The most convenient way to describe the mirror transform is to return
to the Type IIA brane set-up. For now let us ignore the NS5-branes and
concentrate on the $N$ D4-branes alone. We also suppose that the
D4-branes are all coincident.
We now compactify $x^3$ on a circle of
radius $R'$, to distinguish it from the radius of $x^{10}$ defined
earlier. The theory on the $N$ D4-branes is effectively three-dimensional in
the infra-red. The low energy degrees-of-freedom include the Wilson line
around $x^3$ which is an adjoint-valued scalar field that generically
breaks the $\U(N)$
gauge symmetry of the three-dimensional gauge theory to
$\U(1)^N$ via the Higgs mechanism. We define the
abelian components of the Wilson line as $\phi^a=\int_0^{2\pi R'}A_3^a\,dx^3$,
$a=1\ldots,N$. Large
gauge transformations on the circle imply that the Wilson line is actually
a periodic variable
\EQ{
\phi^a\thicksim\phi^a+2\pi n^a\ ,\qquad n^a\in{\mathbb Z}\ .
}
In additional to the Wilson line, is the
three-dimensional $\U(1)^N$ gauge field.
The bosonic part of the classical effective three-dimensional action is
\EQ{
S_{\rm cl}=\frac{2\pi R'}{g^2}\int d^3x\bigg\{\frac1{4\pi^2R^{\prime2}
}\big(\partial_i\phi^a)^2-\tfrac12\big(F_{ij}^a\big)^2
-\frac{i\theta}{8\pi^2}\epsilon_{ijk}\,
\partial_i\phi^aF_{jk}^a\bigg\}\ .
}
Here, $i,j,k=0,1,2$ and we assume that the adjoint scalar in the
four-dimensional theory has
no VEV. The gauge fields may be exchanged
under $S$-duality for $N$ real scalars $\sigma^a$, the dual photons. To
do this one adds a new term to the action
\EQ{
-\frac{i}{2\pi}\int d^3x\,\epsilon_{ijk}\,\partial_i\sigma^aF_{jk}^a\ ,
}
involving $\sigma^a$ which
acts as Lagrange multipliers for the Bianchi identity. The abelian
field strength $F_{ij}^a$ can be now integrated out as a Gaussian
field to give
\EQ{
S_{\rm cl}=\frac{1}{2\pi R'}\int d^3x\bigg\{\frac1{g^2}
\big(\partial_i\phi^a)^2+\frac{g^2}{16\pi^2}\Big(\partial_i\sigma^a+
\frac{\theta}{2\pi}
\partial_i\phi^a\Big)^2\bigg\}\ .
\label{trr}
}
Due to the quantization of magnetic charge,
the dual photons are also periodic variables:
\EQ{
\sigma^a\thicksim\sigma^a+2\pi m^a\ ,\qquad m^a\in{\mathbb Z}\ .
}
The classical effective theory has the form of a sigma model in the
$N$ complex scalar fields
\EQ{
X_a=i\big(\sigma^a+\tau\phi^a\big)\ ,
\label{dxa}
}
which are valued on the
torus that we defined earlier $E_\tau$ with periods
$2\omega_1\equiv2\pi i$ and $2\omega_2\equiv2\pi i\tau$. The
pairing of the Wilson line and dual photons is natural when one
includes supersymmetry since they form the scalar component of a
superfield. The effective three-dimensional theory,
including quantum corrections, is a sigma model involving
$N$ chiral superfields whose scalar components are the fields $X_a$.

This story can be given a brane
interpretation in the following
way. We start with the D4-branes in Type IIA wrapped around $x^3$ with
radius $R'$ and $x^6$ with radius $L$. For small
radius $R'$,
we can now perform a T-duality in $x^3$ to yield the Type IIB
configuration of D3-branes spanning $x^0,x^1,x^2,x^6$. Under this
duality, the string coupling is transformed to $g'_s=g_s\sqrt{\alpha'}/R'$.
We follow this
with an $S$-duality on the four-dimensional theory on the
D3-branes. Under this duality, the three-dimensional effective abelian
gauge field 
is exchanged with the Wilson line of the dual gauge field
around $x^6$. This means that the dual photons are
\EQ{
\sigma^a=\int_0^{2\pi L}\tilde A_6^a\,dx^6\ ,
\label{klob}
}
where $\tilde A$ is the dual gauge field of the D3-branes.
Finally, we perform, once again, a T-duality in $x^3$ to return to a
Type IIA configuration with D4-branes spanning
$x^0,x^1,x^2,x^3,x^6$. However, due to the intervening $S$-duality, the
radius of the $x^3$ is not returned to its original value. The new
radius is $R'g'_s=g_s\sqrt{\alpha'}$. In other words, it is independent
of the radius $R'$.\footnote{All memory of $R'$ is not lost because
the string coupling in the dual theory is $R^{\prime2}/(g_s\alpha')$.}
The Wilson line $\phi^a$ in the original theory is now identified with
the Wilson line of the dual gauge field component $\tilde A_3$ around
the dual $x^3$ circle:
\EQ{
\phi^a=\int_0^{2\pi R'}A_3^a\,dx^3\equiv\int_0^{2\pi
g_s\sqrt{\alpha'}}\tilde A_3^a\,dx^3\ .
\label{kloa}
}
The theory describing the collective dynamics
of these D4-branes is the mirror dual, or ``magnetic'', theory. It is a
five-dimensional theory compactified on ${\mathbb R}^3\times T^2$.
The most significant fact is that the torus
$T^2$ has complex structure $\tau$ and,
up to an overall rescaling, is therefore identified with $E_\tau$.
At low energies the effective three-dimensional theory is described
by the Wilson lines around the two cycles of the torus which are
identified with the Wilson line and dual photons, \eqref{kloa} and
\eqref{klob}, respectively.

The discussion so far has been simplified because we have ignored the
fact that there are NS5-branes in the original Type IIA set-up on
which the D4-branes can fractionate. The gauge group of the effective low
theory should be $\U(1)^n$ (recall $n=kN-k+1$) rather
than $\U(1)^N$. So there should be $n$ complex fields associated to
$n$ Wilson lines and $n$ dual photons. Furthermore, these complex
fields will be valued on tori which are not simply copies of $E_\tau$
but have ``renormalized'' complex structures. This is clear when
one looks at the theory for large $R'$. In that limit, one can think
of the problem in two stages. First, there is an effective
four-dimensional theory {\it \`a la\/} Seiberg and Witten. This is a
$\U(1)^n$ gauge theory with coupling constants encoded in the period
matrix $\tau_{uv}$ of $\Sigma$. The way that all these extra Wilson
lines and dual photons arise in the dual theory
will emerge in due course. For now we must put back the
NS5-branes.
Under the first T-duality the NS5-branes become Type IIB NS5-branes. Then
under $S$-duality they becomes D5'-branes (to distinguish them from
the other D-branes in the problem). Finally the T-duality around
$x^3$ changes them into D4'-branes spanning
$x^0,x^1,x^2,x^4,x^5$, but localized at points on the
$x^3,x^6$ torus. In the original theory, the NS5-branes allow the
D4-branes to fractionate and the gauge group is enhanced from $\U(N)$ to
$\SU(N)^k\times\U(1)$. In the dual magnetic theory, the freedom for
the branes to fractionate is encoded by presence of VEVs for the
impurities. The impurities somehow must also
encode all the extra Wilson lines and
dual photons that we expect, as we shall see later.
As usual in a mirror transform we have mapped
the Coulomb branch of the original theory,
where the D4-branes were prevented from moving off the NS5-branes, to
the Higgs branch of the magnetic theory, where the impurities gain a
VEV Higgsing the dual $\U(N)$ gauge group.

The configuration that we are considering preserves eight real
supersymmetries. So we have a realization of $\ms$ as the Higgs
branch of an $\U(N)$ impurity gauge theory with eight real
supercharges. This is why the mirror map is a useful devise. The Higgs branch
will not be subject to quantum corrections and in this way we are able to
``solve'' the theory. It is naturally described by a set of
$D$-flatness equations which involve the, suitably normalized,
components of the dual $\U(N)$
gauge field $\tilde A_{z,\bar z}=\tfrac12(\tilde A_3\pm i\tilde A_6)$
of the D4-branes along the torus\footnote{As
previously we perform an overall re-scaling of the torus $T^2$ so that it
becomes parameterized by the holomorphic
coordinate $z$ periodic on $\Gamma=2\omega_1{\mathbb
Z}\oplus2\omega_2{\mathbb Z}$.} and the adjoint-valued
complex scalar field $\phi$,\footnote{Not to be confused with the
Wilson line $\phi^a$.} describing the fluctuations of
the D4-branes in the $x^4,x^5$ direction. This field is directly
related to the M-theory picture described in the previous section:
the $N$ eigenvalues of $\phi$ are precisely the $v_a$.
In addition, the D4'-brane impurities
give rise to $k$ hypermultiplets $(Q_i\,\tilde Q_i)$ transforming in the
$(\BN,\overline{\BN})$-representation of $\U(N)$, which are {\it localized\/}
on the torus at points $z_i$, $i=1,\ldots,k$. The $D$-flatness
conditions, with some convenient
choice of normalization of the hypermultiplets, read
\AL{
&\tilde F_{z\bar
z}-[\phi,\phi^\dagger]=-2\pi i\sum_{i=1}^k\delta^2(z-z_i)
(Q_iQ_i^\dagger-\tilde Q^\dagger_i\tilde Q_i)\ ,\label{mma}\\
&\tilde{\cal D}_{\bar z}\phi=2\pi i\sum_{i=1}^k\delta^2(z-z_i)Q_i\tilde Q_i\ .
\label{mmb}
}
Here, \eqref{mma} is a real equation and \eqref{mmb} is a complex
equation and $\tilde {\cal D}_{\bar z}\phi=\partial_{\bar
z}\phi+[\tilde A_{\bar
z},\phi]$. These equations are a generalization of Hitchin's
self-duality equations reduced to two dimensions \cite{hitchin}.

The space $\ms$ described as the solution to \eqref{mma}-\eqref{mmb}
modulo $\U(N)$ gauge transformations, has another interesting interpretation.
Consider a configuration of $N$ D2-branes lying inside $k$ coincident
D6-branes. The
D2-branes are embedded as a charge $N$ Yang-Mills instanton solution
in the $\U(k)$ gauge theory on the D6-branes
in the four directions orthogonal to the D2-branes inside the D6-branes.
In the brane picture, the moduli space of the instantons is described by the
Higgs branch of an $\U(N)$ gauge theory, with eight
real supercharges and $k$ hypermultiplets transforming in the
$(\BN,\overline{\BN})$. This yields nothing but the ADHM construction of
instantons in ${\mathbb R}^4$ $\U(k)$ gauge theory.
Now imagine compactifying two spatial
dimensions in the D6-branes orthogonal to the D2-branes.
The D2-branes now correspond to Yang-Mills instantons in a $\U(k)$ gauge
theory on ${\mathbb R}^2\times
T^2$ rather than ${\mathbb R}^4$.
Performing a $T$-duality along each compactified direction, we now
have the configuration of $N$ D4-branes and $k$ D4'-branes described above.
So the Coulomb branch of the compactified quiver theory is identified with
the moduli space of $N$ instantons in $\U(k)$ gauge theory on
${\mathbb R}^2\times T^2$.

{\it A hyper-K\"ahler quotient}

Lurking within the generalized Hitchin equations \eqref{mma}-\eqref{mmb}
is a complexified classical integrable
dynamical system and uncovering it will be key to our story of the
superpotential of the three-dimensional theory. But this is jumping
ahead. The first thing to say about these Hitchin's equations is
they are an example
of a hyper-K\"ahler quotient, although of a rather novel
infinite-dimensional kind. The quotient construction \cite{HKQ}
provides a way of
constructing a new hyper-K\"ahler manifold $\ms$ in terms of an old
one $\tilde\ms$ with some group of isometries $G$. These isometries must
preserve both the metric and and the three complex structures and so
are described by {\it tri-holomorphic\/} Killing
vectors $X_r$, $r=1,\ldots,{\rm dim}\,G$. Each Killing vector defines
a triplet of moment maps $\vec\mu^{X_r}$ defined by
$d\vec\mu^{X_r}=i(X_r)\vec\omega$, where $\vec\omega$ are the triplet of
K\"ahler forms. The quotient space is defined as the coset manifold
$\ms=\ns/G$, where the level set $\ns\subset\tilde\ms$ is defined by
$(\vec\mu^{X_r})^{-1}(0)$, for all $r$, {\it i.e.\/}~the subspace of $\tilde\ms$ on
which the moment maps vanish. It can be shown that $\ms$
inherits the hyper-K\"ahler structure of $\tilde\ms$. If $\tilde\ms$ has real
dimension $4n$ then the above construction leads to $\ms$ with
dimension $4(n-{\rm dim}\,G)$.

In the present context, the novelty arises from the fact that the
space $\ms$, parameterized by $\phi$, $\tilde A$, $Q$ and $\tilde
Q$, is infinite dimensional because $\phi$ and the gauge field $\tilde
A$ are
{\it functions\/} on $E_{\tau}$. However, the quotient construction---at least
formally---proceeds in the usual way. In this case the quotient group
$G$ is the group of $\U(N)$ local gauge transformations on $E_{\tau}$. The idea
is that quotienting the infinite-dimensional space $\tilde\ms$ by the
infinite-dimensional gauge group leads to a
finite-dimensional space $\ms$. The equations \eqref{mma} and
\eqref{mmb} are really a triplet of equations since \eqref{mmb} is a
complex equation while \eqref{mma} is real, and they
are nothing but the moment maps for the local $\U(N)$ action.

But, as pointed by Kapustin \cite{Kapustin:1998xn}, there is a problem.
The fields $\phi$ and $\tilde A$ must have
simple poles at the punctures $z_i$ with residues that are variable,
given as they are by the $Q$ and $\tilde Q$. This means that the
variations of $\phi$ and $\tilde A$ will have simple poles as well and so
their norm will be logarithmically divergent.
As a consequence, the corresponding tangent vectors to $\ms$ will
have infinite norm. Physically this is a different manifestation of the
fact that in the Type IIA brane set-up the relative centres-of-mass of
the stacks of D4-branes are not genuine moduli. These parameters are, of
course, the masses $m_i$ of the bi-fundamental hypermultiplets. All
this means is that we have not quite described the correct hyper-K\"ahler
quotient. What we need to do is freeze the impurities $Q$ and $\tilde
Q$ in some way to ensure that the non-normalizable modes are
discarded. But we must do this in a way which still leads to a
hyper-K\"ahler manifold. The way to do this was described by Kapustin
\cite{Kapustin:1998xn}, however, we will present a re-interpretation of his
procedure which is more in keeping with the quotient construction.

We now hypothesize that the correct construction involves a
hyper-K\"ahler quotient by a larger group which involves not only
local $\U(N)$ gauge transformations on the torus $E_{\tau}$, but also
$\U(1)$ transformations at each puncture:
\EQ{
Q_{ai}\to Q_{ai}e^{\psi_i}\ ,\quad \tilde Q_{ia}\to e^{-\psi_i}\tilde
Q_{ia}\ .
\label{wee}
}
In fact the additional part of the quotient group is $\U(1)^{k-1}$
because the transformation that rotates all the $Q$ and $\tilde Q$ by
the same phase is part of the $\U(N)$ gauge group.
The associated moment maps are simply
\AL{
&\sum_a\big(Q_{ai}Q^\dagger_{ia}-\tilde Q^\dagger_{ai}\tilde
Q_{ia}\big)=0\ ,\label{ffa}\\
&\sum_aQ_{ai}\tilde Q_{ia}=0\ .
\label{ffb}
}
The question is where are the mass parameters? In the hyper-K\"ahler
quotient construction, the definition of the moment maps for any
abelian components of $G$ is ambiguous. Any constant can be
added. When the hyper-K\"ahler quotient construction is viewed as the
Higgs branch of a supersymmetric gauge theory with eight supercharges,
these parameters are the Fayet-Illiopolos (FI) couplings of the
abelian part of the gauge group. In the present context, the masses $m_i$
are the complex FI couplings
associated to the abelian subgroup $\U(1)^k$ which acts as in \eqref{wee}.
The complex components of the moment maps
\eqref{mmb} and \eqref{ffb} are modified to
\AL{
&\tilde {\cal D}_{\bar z}\phi=2\pi i\sum_{i=1}^k\delta^2(z-z_i)\Big(
Q_i\tilde Q_i-\frac m{k}1_{\sst[N]\times[N]}\Big)\ ,\label{ssa}\\
&
\sum_aQ_{ai}\tilde Q_{ia}=Nm_i\ .
\label{ssb}
}
As before $m=\sum_im_i$.
The form of \eqref{ssa} is consistent with \eqref{ssb} in the
following way. The quantity $\varphi={\rm Tr}\phi$ satisfies
\EQ{
\partial_{\bar z}\varphi=2\pi iN\sum_{i=1}^k\delta^2(z-z_i)\Big(
m_i-\frac m{k}\Big)\ ;
}
and hence, is a meromorphic function
on $E_{\tau}$ with simple poles at $z=z_i$ and residues $N(m_i-m/k)$. For
consistency the sum of the residues must vanish and this is guaranteed
by \eqref{gmass}.

We can now go on and find a very concrete realization of the
quotient. Firstly, as we have already mentioned,
the real and complex moment maps, \eqref{mma} and
\eqref{ffa} verses \eqref{ssa} and \eqref{ssb}, can be viewed as the
$D$- and $F$-flatness conditions of a Higgs branch of supersymmetric gauge
theory with four real supercharges.
As usual, we can relax the $D$-flatness
condition and simply impose the $F$-flatness conditions
\eqref{ssa}-\eqref{ssb} at the expense of modding out by the
complexified gauge group. Picking out the complex moment map amounts
to picking out a preferred complex structure from the three
independent complex structures of the hyper-K\"ahler manifold. The
associated preferred K\"ahler form will turn out to be the symplectic
form of the complexified dynamical system.

It is convenient to introduce the ``spins''
$\S^i$, $N\times N$ matrices at each puncture, with elements
\EQ{
\S^i_{ab}\ \overset{\rm def}=\ Q_{ai}\tilde Q_{ib}-\frac{m}{k}\delta_{ab}\ .
\label{fsp}
}
The complex moment map equations \eqref{ssa}-\eqref{ssb} are then
\AL{
\tilde {\cal D}_{\bar z}\phi&=2\pi i\sum_{i=1}^k\S^i\delta^2(z-z_i)\ ,
\label{poi}\\
\sum_a\S^i_{aa}&=N\big(m_i-\frac mk\big)\ .
\label{zzb}
}
The constraint \eqref{zzb}
is equivalent to Kapustin's residue condition, generalizing the
residue condition of Donagi and Witten \cite{donwitt}, because it implies
that the spins lie in a particular conjugacy class of the complexified
quotient group:
\EQ{
\S^i=U_i\begin{pmatrix}Nm_i-\tfrac m{k}&&&\\ &-\tfrac m{k}&&\\
&&\ddots&\\ &&&-\tfrac m{k}\end{pmatrix}
U^{-1}_i
}
for elements $U_i$ in the complexified quotient group.
\footnote{To see this identify $(U_i)_{a1}=Q_{ai}/\sqrt{Nm_i}$ and
$(U^{-1}_i)_{1a}=\tilde{Q}_{ia}/\sqrt{Nm_i}$. 
The fact that $\sum_a(U_i^{-1})_{1a}(U_i)_{a1}=1$ follows
from \eqref{ssb}.}

To proceed, it is very convenient to use up (most of) the local part
of the quotient group, $\GL(N,{\mathbb C})$, to transform the
the anti-holomorphic component $A_{\bar
z}$ into a constant diagonal matrix:
\EQ{
\tilde A_{\bar z}=\frac{\pi i}{2(\bar\omega_2\omega_1-\bar\omega_1\omega_2)}
{\rm diag}(X_1,\ldots,X_N)=-\frac{1}{16g^2}{\rm
diag}(X_1,\ldots,X_N)\ .
}
The only local transformations that remain act by shifting the
$X_a$ by periods of $E_{\tau}$,
\EQ{
X_a\to X_a+2n\omega_1+2m\omega_2\ ,\qquad m,n\in{\mathbb Z}\ .
}
The $X_a$ are nothing but the complex combination of the abelian Wilson lines
and dual photons defined in \eqref{dxa}.
The global part of the gauge group is also fixed, up to
permutations of the $X_a$ and the $\GL(1,{\mathbb C})^N$ diagonal
subgroup, as well as the complexified $\GL(1,{\mathbb C})^{k-1}$
``flavour'' symmetries \eqref{wee}.
Putting these symmetries together, we must mod
out by the action
\EQ{
Q_{ai}\to \zeta_aQ_{ai}\xi_i\ ,\qquad\tilde Q_{ia}\to\xi_i^{-1}\tilde
Q_{ia}\zeta_a^{-1}\ .
\label{llw}
}

We can now solve explicitly for $\phi$ to get a very concrete parameterization
of the quotient space $\ms$ and the associated dynamical system.
Roughly
speaking, the elements of $\phi$ must have simple poles at $z=z_i$ to
account for the $\delta$-functions. Candidate functions are
$\zeta(z-z_i)$ and $\sigma(z-z_i)^{-1}$,\footnote{A short review of
elliptic functions and their properties is provided in Appendix A.}
however, these must be put
together in the right way to ensure periodicity on $E_{\tau}$. After
some trial and error, one is led to the solution
\EQ{
\phi_{aa}(z)=p_a+\sum_{i=1}^k\S^i_{aa}\zeta(z-z_i)\ ,
\label{ggf}
}
for the diagonal components, where the $p_a$ are new parameters. There
are extra constraints on the spins, arising from the fact that in the
gauge we have chosen, the diagonal elements $\phi_{aa}(z)$ are
meromorphic functions on $E_{\tau}$ and so the sum of the residues
must vanish:
\EQ{
\sum_i\S^i_{aa}=0\ .
\label{zza}
}
The off-diagonal elements are
\EQ{
\phi_{ab}(z,\bar z)=e^{\psi(z,\bar z)X_{ab}}\sum_{i=1}^k \S^i_{ab}
\frac{\sigma(X_{ab}+z-z_i)}{\sigma(X_{ab})\sigma(z-z_i)}e^{-\psi(z_i,\bar
z_i)X_{ab}}
\qquad(a\neq b)\ .
\label{ggg}
}
Here $X_{ab}\equiv X_a-X_b$, and we have
defined
\EQ{
\psi(z,\bar z)\ \overset{\rm def}=\
\frac1{\bar \omega_2\omega_1-\bar
\omega_1\omega_2}\big[\zeta(\omega_2)
(\bar \omega_1 z-\omega_1\bar z)-\zeta(\omega_1)(\bar
\omega_2z-\omega_2\bar z)\big]\ .
}
One can readily verify that $\phi_{ab}(z,\bar z)$ is periodic on
$E_\tau$. Furthermore, a shift of $X_a$ by a lattice vector
$2\omega_\ell$, can be undone by a large gauge transformation on the
torus as anticipated earlier.

We now have an explicit
parameterization of $\ms$ furnished by $p_a$, $\S^i_{ab}$ and
$X_a$. In order to determine the dimension of $\ms$ we now count the
number parameters. First of all the spins. There are
$4Nk$ real quantities $Q$ and $\tilde Q$ subject to
$2k$ and $2N$ real constraints, \eqref{zza} and \eqref{zzb},
respectively. The group action \eqref{llw} further reduces the number
of variables by $2(N+k-1)$ (the ``1'' arising from the fact that not
all the parameters $\zeta_a$ and $\xi_i$ in \eqref{llw} are
independent). Hence the number of real independent spin variables is
$4(Nk-N-k+1)$. The remaining variables are $p_a$ and $X_a$, giving
$4N$ real parameters. Hence, the overall dimension
of the quotient space $\ms$ is $4n\equiv 4(kN-k+1)$. From the point
of view of the dynamical system, the variables $p_a$ are
naturally the momenta conjugate
to the $X_a$. Notice that $\phi(z,\bar z)$ is only
dependent on the differences $X_{ab}\equiv X_a-X_b$; hence $\sum_ap_a$
is not dynamical. In fact $\sum_ap_a$, and its conjugate
position $\sum_aX_a$, correspond to the decoupled overall $\U(1)$ factor of the
gauge group in the quiver theory. We now choose to set
\EQ{
\sum_ap_a=\sum_aX_a=0\ .
\label{rouo}
}
Notice that although we have a concrete parameterization of $\ms$, the
relation with the physical parameters of the three-dimensional Coulomb
branch is not obvious. In the basic $\N=1^*$ case, recovered by taking
$k=1$, the spins are completely absent. The coordinates in this case
are $p_a$ and $X_a$. Writing $X_a=-i(\sigma^a+\tau\phi^a)$,
the real components $\sigma^a$ and
$\phi^a$ are precisely the dual photons and Wilson lines of the
effective $\U(1)^N$ theory. In the general case, the relation with the
dual photons and Wilson lines and the
coordinates is less obvious. Intuitively, the $X_a$ are dual photons
and Wilson lines of the diagonally embedded $\SU(N)$.

\subsection{A dynamical system and the exact superpotential}

As we have already alluded to above,
there is also a completely integrable dynamical
system underlying the construction of $\ms$, for which $\ms$ is the
phase space with symplectic form given by the K\"ahler form singled out
by the complex moment map.
It is the rather esoteric spin generalization of the
elliptic Calogero-Moser system which was first described in Ref.~\cite{GH}
and further studied in Refs.~\cite{Krichever:1994vg,Nekrasov:1996nq}.
The integrable system has an associated {\it spectral curve\/} which
is defined by \cite{Krichever:1994vg}
\EQ{
F(z,v)={\rm det}\,\big(v1_{\sst[N]\times[N]}-\phi\big)=0\ .
}
This is precisely the Seiberg-Witten curve $\Sigma$, the branched
$N$-fold cover of $E_\tau$ which appeared in the M-theory
interpretation, Eq.~\eqref{curve}. Since
the dynamical system is completely integrable, there are $n$ (complex)
Hamiltonians. These are identified with coordinates on the
Coulomb branch of the four-dimensional theory.
The conjugate angle variables (also complex),
$X_u$, $u=1\ldots,n$, take values in the Jacobian variety
${\EuScript J}(\Sigma)$. Finally we have identified all
the Wilson lines
and dual photons of the three-dimensional effective theory. The
explicit maps between the variables $\{p_a,X_a,\S^i\}$ and the angle
variables was found in \cite{Krichever:1994vg,Nekrasov:1996nq}, but we
shall not need them here.

Before proceeding, the resulting equations are cleaner if
the spin variables are re-defined by
\EQ{
\S^i_{ab}\to \S^i_{ab}e^{\psi(z_i,\bar z_i)X_{ab}}\ .
}
The Hamiltonians of the dynamical system arise as the residues of the
gauge invariant quantities ${\rm Tr}\phi^l(z)$, $l=1,2,\ldots,N$. Since
the system is integrable, there will be $2(kN-k+1)$ independent
Hamiltonians which will be identified with coordinates on the Coulomb
branch of the four-dimensional $\N=2$ theory that we started
with. Of particular importance will be the $k$ combinations of
Hamiltonians that correspond
to the condensates $u_2^{(i)}=\langle{\rm Tr}\,\Phi_i^2\rangle$ for each of the
$\SU(N)$ factors of the gauge group. They must come,
on dimensional grounds,
from expressions quadratic in $\phi(z)$. There are two such terms
$\big({\rm Tr}\,\phi(z)\big)^2$ and ${\rm Tr}\,\phi(z)^2$. We have already
argued that the quantity that relates to the $\N=1^*$ deformation
is $u_2(z)=\tfrac1{2N}\sum_{a\neq b}(v_a(z)-v_b(z))^2$, and since
$v_a(z)$ are the eigenvalues of $\phi(z)$, this identifies
\EQ{
u_2(z)={\rm Tr}\,\phi^2-\tfrac1N\big({\rm Tr}\,\phi\big)^2\ .
}
We now compute this quantity given our solution for $\phi(z)$ in
\eqref{ggf} and \eqref{ggg}.

Since $\phi(z)$ has simple poles, the quadratic invariant $u_2(z)$
has double poles at $z_i$. Since it is manifestly
elliptic, the expansion must be of the form (compare \eqref{u2exp})
\EQ{
u_2(z)=\sum_{i=1}^k\lambda_i\wp(z-z_i)+\sum_{i=1}^k\zeta(z-z_i)H_i+H_0\ ,
\label{expan}
}
where $\sum_{i=1}^kH_i=0$. It is tedious but a straightforward
exercise to extract
the residues and constant part. Firstly, the residues of the double
poles are constants
\EQ{
\lambda_i=\sum_{ab}\S^i_{ab}\S^i_{ba}-\tfrac1N(\sum_a\S^i_{aa})^2=
N(N-1)m_i^2\ .
}
The residues of the simple poles are non-trivial functions on $\ms$, to wit
\SP{
H_i&=2\sum_ap_a\S^i_{aa}
-2N\big(m_i-\tfrac mk\big)\sum_{j(\neq i)}\big(m_j-\tfrac
mk\big)\zeta(z_{ij})\\
&+2\sum_a\sum_{j(\neq
i)}\S^i_{aa}\S^j_{aa}\zeta(z_{ij})-2\sum_{a\neq b}\sum_{j(\neq
i)}\S^i_{ab}\S^j_{ba}\frac{\sigma(X_{ab}+z_{ji})}{\sigma(X_{ab})
\sigma(z_{ji})}\ ,
\label{dhi}
}
where $z_{ij}\equiv z_i-z_j$. One can verify that $\sum_iH_i=0$.
To complete the expansion,
the constant piece in the expansion is
\SP{
H_0&=\sum_ap_a^2-\sum_{a\neq
b}\sum_i\S^i_{ab}\S^i_{ba}\wp(X_{ab})\\
&+\sum_{a\neq b}\sum_{i\neq
j}\S^i_{ab}\S^j_{ba}\frac{\sigma(X_{ab}+z_{ji})}
{\sigma(X_{ab})\sigma(z_{ji})}\big(\zeta(X_{ab}+z_{ji})-\zeta(X_{ab})\big)\\
&-\tfrac12\sum_{i\neq
j}\Big[\sum_a\S^i_{aa}\S^j_{aa}-N
\big(m_i-\tfrac mk\big)\big(m_j-\tfrac mk\big)\Big]
\Big(\wp(z_{ij})-\zeta(z_{ij})^2\Big)\ .
\label{dhz}
}
The $k$ independent functions
on $\ms$, $H_0$ and $H_i$, are a subset of the Hamiltonians of the
integrable system.

The Hamiltonians are not
simply invariant under these shifts \eqref{newduality}. Rather
the shifts can be undone by an appropriate
transformation on $\ms$. To find this transformation, under a shift of
$z_l\to z_l+2\omega_\ell$, we have
\EQ{
H_i\big(p_a,X_a,\S^j_{ab}
\big|z_j+2\delta_{jl}\omega_\ell\big)=H_i\big(p'_a,X_a,
\S^{\prime j}_{ab}\big|z_j\big)\ ,
\label{gyyt}
}
where
\EQ{
p'_a=p_a-2\zeta(\omega_\ell)\big(\S^l_{aa}-2(m_i-\tfrac
mk)\big)\ ,\qquad \S^{\prime j}_{ab}=\S^j_{ab}
e^{2\delta_{jl}X_{ab}\zeta(\omega_\ell)}\ .
\label{trans}
}
The remaining Hamiltonian transforms similarly, but with an additive
anomaly
\EQ{
H_0\big(p_a,X_a,\S^j_{ab}\big|z_j+2\delta_{jl}\omega_\ell
\big)=H_0\big(p'_a,X_a,
\S^{\prime j}_{ab}\big|z_j\big)+2\zeta(\omega_\ell)H_l\big(p'_a,X_a,
\S^{\prime j}_{ab}\big|z_j\big)\ .
\label{yyt}
}

We have already described, based on the structure of the massive vacua
in the M Theory picture, how to relate the quantities $H_i$ and $H_0$,
now interpreted as Hamiltonians, to the condensates. The conclusion
was that a basis of functions with the right modular properties is
provided by the $H_i$ and $H^*$ defined in \eqref{defhstar}. But, using the
parameterization of the Coulomb branch provided by the integrable
system, we can now see that the basis $\{H_i,H^*\}$ is precisely the
one with good modular properties, not just at the massive vacua, but
also at generic points on the Coulomb branch. To see this, notice that
under shifts of $z_i$ on $E_\tau$, $H^*$ transforms covariantly as the $H_i$
\eqref{gyyt} since the anomaly piece in \eqref{yyt} is compensated.
Based on the semi-classical limit,
the quantity $H^*$ was then identified with the diagonal combination.
Explicitly
\SP{
H^*&=\sum_ap_a^2-\sum_{a\neq
b}\sum_i\S^i_{ab}\S^i_{ba}\wp(X_{ab})
-\tfrac2k\sum_{i\neq l}\bigg\{\sum_ap_a\S^i_{aa}\\
&-N\big(m_i-\tfrac mk\big)\sum_{j(\neq i)}\big(m_j-\tfrac
mk\big)\zeta(z_{ij})+\sum_a\sum_{j(\neq
i)}\S^i_{aa}\S^j_{aa}\zeta(z_{ij})\bigg\}\zeta(z_{il})\\
&+\sum_{a\neq b}\sum_{i\neq
j}\S^i_{ab}\S^j_{ba}\frac{\sigma(X_{ab}+z_{ji})}
{\sigma(X_{ab})\sigma(z_{ji})}\Big(\zeta(X_{ab}+z_{ji})-\zeta(X_{ab})
+\tfrac2k\sum_{l(\neq i)}\zeta(z_{il})
\Big)\\
&-\tfrac12\sum_{i\neq
j}\bigg\{\sum_a\S^i_{aa}\S^j_{aa}-N
\big(m_i-\tfrac mk\big)\big(m_j-\tfrac mk\big)\bigg\}
\Big(\wp(z_{ij})-\zeta(z_{ij})^2\Big)\ .
\label{spot}
}
The exact superpotential of the three-dimensional theory
corresponding to the diagonal $\N=1^*$
deformation is then simply $W=-k\mu H^*/g^2$.

{\it The massive vacua}

The superpotential $W=-k\mu\,H^*/g^2$ determines the vacuum structure of the
$\N=1^*$ deformation of the theory. A full
analysis of the vacuum structure is beyond the scope of the present
work. Actually, even in the $\N=4$ case, where the
superpotential is a good deal simpler, there is only a systematic
treatment of the massive vacua for $N\geq3$ \cite{Nick}.
We shall achieve as much for the finite $\N=2$ theories.

The massive
vacua have a very beautiful interpretation from the point-of-view of
the dynamical system: they are precisely {\it equilibrium
configurations\/} with respect to the space of flows defined by
the $n$ Hamiltonians.\footnote{Here, ``time'' is an
auxiliary concept referring to
evolution in the dynamical system and not a spacetime concept in the
field theories under consideration.}
To see this, recall that the massive vacua correspond to points of the
four-dimensional Coulomb branch for which $\Sigma$ degenerates to a
torus: cycles pinch off and one is
left with an $N$-fold un-branched cover of $E_\tau$. This means that the
Jacobian Variety ${\EuScript J}(\Sigma)$ itself
degenerates: at these points the period matrix only has rank 1, with
non-zero eigenvalue $\tau$. The remaining torus is associated with the
overall $\U(1)$ factor of the gauge group which we have removed from
the integrable system \eqref{rouo}.
So at a massive vacuum, the remaining angle variables must stay fixed
under any time evolution. Since the Hamiltonians are by
definition constants of the motion, this means that the entire
dynamical system must be static at a massive vacuum and the system is
at an equilibrium point. Consequently, a
massive vacuum is not only a critical point of the Hamiltonian
describing the $\N=1^*$ deformation, but
simultaneously of all the other $n-1$ Hamiltonians. On the other hand,
for the massless vacua, this is no longer true.

First of all, it is instructive to recall some details of the
$\N=4$ case described in \cite{Nick}
which is recovered in our formalism by setting $k=1$.
This will provide an important clue for quiver theories. When
$k=1$, the constraints on the single spin mean that it is not dynamical:
\EQ{
k=1:\qquad \S_{ab}=m\Big(1-\delta_{ab}\Big)\ .
}
There is a single quadratic Hamiltonian,
\EQ{
k=1:\qquad H_0=\sum_ap_a^2-m^2
\sum_{a\neq
b}\wp(X_{ab})\ .
}
At the critical points of $H_0$ the momenta $p_a$ conjugate to $X_a$
vanish. The
massive vacua are associated to the finer lattices
$\Gamma'$ which contain
$\Gamma$ as a sublattice of index $N$. This means that there are $N$
points of $\Gamma'$ contained in a period parallelogram of $\Gamma$.
The simplest kind, labelled by two integers $p$ and $q$
with $pq=N$, are generated by $2\omega_1/q$ and $2\omega_2/p$. All the
other cases can be generated from these by modular transformations,
as we shall see later. Each
$a\in\{1,\ldots,N\}$ is uniquely associated to the pair $(r_a,s_a)$, with
$0\leq r_a<q$ and $0\leq s_a<p$. The critical point of $H_0$
associated to $(q,p)$ is then
\EQ{
X_a=\frac{2r_a}{q}\omega_1+\frac{2s_a}{p}\omega_2\qquad 0\leq r<q,\ 0\leq s<p\
.
\label{bcase}
}
The proof that this is a critical point of $H_0$ is delightfully
simple. One only needs to use the fact that $\wp'(z)$ is an
odd elliptic function. Terms in the
sum $\sum_{b(\neq a)}\wp'(X_{ab})$ either cancel in pairs or vanish
because $X_{ab}$ is a half-lattice point. As we mentioned, the set
\eqref{bcase} does not exhaust the set of massive vacua.
For a given pair
$(q,p)$ we can generate $p-1$ additional vacua by performing the
suitable modular transformation on $\tau$ to give
\EQ{
X_a=\frac{2r_a}{q}\omega_1+
\frac{2s_a}{p}\Big(\omega_2+\frac lq\omega_1\Big)\qquad
0\leq l<p\ .
\label{nvc}
}
So the total number of massive vacua is equal to $\sum_{p|N}p$, as
expected on the basis of the semi-classical analysis.

Returning to the finite $\N=2$ theories, we hypothesize that the
massive vacua are also associated to the finer lattices $\Gamma'$
with $X_a$ given in \eqref{nvc}. Once again we can start with the
configurations \eqref{bcase} and the additional massive
vacua will be obtained by modular transformations. We will now find a
set of critical points common to each of the Hamiltonians $H_i$ and
$H_0$, and so, by implication, of $H^*$. Firstly, by varying
$H_0$ with respect to the $p_a$, we find, as in the $\N=4$
case described above, that $p_a=0$. Varying each $H_i$ with respect to
$p_a$, one finds
\EQ{
\S^i_{aa}=m_i-\frac mk\ .
\label{pce}
}
What remains is to impose the $X_a$ and $\S^i_{ab}$ (more
properly the $Q_{ai}$ and $\tilde Q_{ia}$)
equations and hence find the spins at the critical points. Rather than
write down these equations and find their solutions, which is
necessarily complicated and uninspiring, we can motivate the form of
the solution and then verify that the ansatz is a critical point.
The critical point equations are significantly
simplified if, for $a\neq b$,
\EQ{
\S^i_{ab}\S^j_{ba}\frac{\sigma(X_{ab}+z_{ji})}{\sigma(X_{ab})\sigma(z_{ji})}
\label{blob}
}
with $X_a$ as in \eqref{bcase}, are periodic in the indices $r_a$ (mod
$q$) and $s_a$ (mod $p$). This can be achieved if
\EQ{
\S^i_{ab}\thicksim \rho_i^{r_a-r_b}\lambda_i^{s_a-s_b}
e^{2z_i\big[\frac{r_a-r_b}q\zeta(\omega_1)+\frac{s_a-s_b}p
\zeta(\omega_2)\big]}\ .
\label{pop}
}
Here, $\rho_i$ and $\lambda_i$ are arbitrary $q^{\rm th}$ and $p^{\rm
th}$ roots of unity, respectively. In fact, this periodicity
requirement determines
the spins completely, up to an overall factor which is easily fixed.
With a little more work one is lead to the ansatz
\EQ{
\S^i_{ab}=m_i\rho_i^{r_a-r_b}\lambda_i^{s_a-s_b}
e^{2z_i\big[\frac{r_a-r_b}q\zeta(\omega_1)+\frac{s_a-s_b}p\zeta(\omega_2)
\big]}-
\frac{m}{k}\delta_{ab}\ .
\label{sols}
}
Notice that this is consistent with \eqref{fsp} where
\EQ{
Q_{ai}=\sqrt{m_i}\rho_i^{r_a}\lambda_i^{s_a}
e^{2z_i\big[\frac{r_a}q\zeta(\omega_1)+\frac{s_a-s}p\zeta(\omega_2)
\big]}\ ,\qquad
\tilde Q_{ia}=\sqrt{m_i}\rho_i^{-r_a}\lambda_i^{-s_a}
e^{-2z_i\big[\frac{r_a}q\zeta(\omega_1)+\frac{s_a}p\zeta(\omega_2)
\big]}\ .
}
In particular, the constraints \eqref{zza} and \eqref{zzb} are
satisfied. In addition, the solution is consistent with \eqref{pce}.
The undetermined roots of unity $\rho_i$ and
$\lambda_j$ label
inequivalent critical points. Hence the number of
critical points appears to be $p^kq^k=N^k$. But this over counts.
The residual $\U(1)^k$
transformations described previously
can be used to set $\rho_i=\lambda_i=1$ for one
particular $1\leq i\leq k$, and so there is an $N^{k-1}$ additional
degeneracy of vacua for each
$N=pq$: precisely the same counting that we found in
\eqref{massivevacua} for the
massive vacua. This gives an important clue that the critical points
we have found are to be identified with
the massive vacua. But there is more. We saw that
the massive vacua were related by $(2\omega_1,2\omega_2)$ translations of the
punctures $z_i$. This should be reflected by the critical points of
the superpotential. Consider the quantity \eqref{blob} with the
spins as in \eqref{sols}. Under the shift $z_i\to
z_i+2m\omega_1+2n\omega_2$, for $a\neq b$,
\EQ{
\S^i_{ab}\S^j_{ba}
\frac{\sigma(X_{ab}+z_{ji})}{\sigma(X_{ab})\sigma(z_{ji})}
\longrightarrow e^{-2\pi im\frac{s_a-s_b}p+2\pi in\frac{r_a-r_b}q}\,
\S^i_{ab}\S^j_{ba}
\frac{\sigma(X_{ab}+z_{ji})}{\sigma(X_{ab})\sigma(z_{ji})}\ .
}
To show this one has to employ the identity
$\omega_2\zeta(\omega_1)-\omega_1\zeta(\omega_2)=i\pi/2$.
Consequently, these
transformations change the roots of unity labelling the vacua as
\SP{
&z_i\to z_i+2\omega_1:\qquad \lambda_i\to \lambda_ie^{-\frac{2\pi
i}p}\ ,\ \rho_i\to\rho_i\ ,\\
&z_i\to z_i+2\omega_2:\qquad \lambda_i\to\lambda_i\ ,\
\rho_i\to \rho_ie^{\frac{2\pi i}q}\ .
\label{zis}
}
Hence, as found earlier, shifts of the punctures by the lattice
$(2\omega_1,2\omega_2)$ does indeed
permute the $N^{k-1}$ massive vacua for a given $pq=N$. Shifts on the
larger lattice, generated by
$(2\tilde\omega_1,2\tilde\omega_2)\equiv(2p\omega_1,2q\omega_2)$,
leave the vacua invariant. (Notice that the relation of $p$ and $q$ to
$\omega_1$ and $\omega_2$ swaps over relative to \eqref{bcase}.)

{\bf PROOF:} We now prove that our ansatz \eqref{bcase} and
\eqref{sols} is a critical point of $H_i$ and $H_0$.
First of all, consider the $X_a$ variations. The $X_a$-derivative of
the second term of $H_0$ in Eq.~\eqref{dhz} is
\EQ{
-2\sum_im_i^2\sum_{b(\neq a)}\wp'(X_{ab})\ .
}
This vanishes for the same reason as in the $k=1$ case.
For each $a$ and $b$ there exists a unique
$b'$ (possibly $b'=b$) such that
\EQ{
X_{ab}=X_{b'a}\ \text{mod}\ \Gamma\ .
\label{dbp}
}
Then we can see that terms cancel in pairs when $b\neq b'$, since
$\wp'(X_{ab})+\wp'(X_{ab'})=0$, or $\wp'(X_{ab})=0$ when $b=b'$, since
$X_{ab}$ is then a half-lattice point and $\wp'(z)$ is an odd elliptic
function.
Now consider the $X_a$-derivatives of the third term of $H_0$,
in \eqref{dhz},
and the fourth term of $H_i$, in \eqref{dhi}. In both cases, the
resulting expression involves terms like
\EQ{
\sum_{b(\neq a)}\bigg\{\S^i_{ab}\S^j_{ba}
\frac{\sigma(X_{ab}+z_{ji})}{\sigma(X_{ab})
\sigma(z_{ji})}f(X_{ab})-
\S^i_{ba}\S^j_{ab}
\frac{\sigma(X_{ba}+z_{ji})}{\sigma(X_{ba})
\sigma(z_{ji})}f(X_{ba})\biggr\}\ ,
\label{xad}
}
for some function $f(X_{ab})$ elliptic in $X_{ab}$.
Using the special periodicity properties of the
quantity \eqref{blob}, one can show
\EQ{
\S^i_{ba}\S^j_{ab}
\frac{\sigma(X_{ba}+z_{ji})}{\sigma(X_{ba})
\sigma(z_{ji})}=\S^i_{ab'}\S^j_{b'a}\frac{\sigma(X_{ab'}+z_{ji})}
{\sigma(X_{ab'})
\sigma(z_{ji})}\ .
}
Furthermore, $f(X_{ba},z_{ji})=f(X_{ab'},z_{ji})$. After re-labelling
the second term in Eq.~\eqref{xad} with $b'\to b$,
we see that the two terms in \eqref{xad} cancel.

Now we turn to the spins. It is convenient to parameterize
\EQ{
Q_{ai}=t_{ai}\ ,\qquad\tilde Q_{ia}=y_{ai}/t_{ai}\ .
}
The constraints \eqref{zza} and \eqref{zzb} are then linear in $y_{ai}$:
\EQ{
\sum_ay_{ai}=Nm_i\ ,\qquad\sum_iy_{ai}=m\ .
\label{coy}
}
The symmetries \eqref{llw} can be used to set $t_{a1}=t_{1i}=1$.
Now consider
\SP{
\frac{\partial H_0}{\partial y_{ai}}
=&-2\sum_{b(\neq a)}y_{bi}\wp(X_{ab})-\sum_{j(\neq
i)}\big(y_{aj}-\tfrac m{k}\big)
\big(\wp(z_{ij})-\zeta(z_{ij})^2\big)\\
&+\frac2{y_{ai}}\sum_{b(\neq a)}\sum_{j(\neq i)}\S^i_{ba}\S^j_{ab}
\frac{\sigma(X_{ba}+z_{ji})}
{\sigma(X_{ba})\sigma(z_{ji})}\big(\zeta(X_{ba}+z_{ji})-\zeta(X_{ba})\big)
+B_i+C_a\ .
\label{mess}
}
Here, $B_i$ and $C_a$ are the Lagrange multipliers for the constraints
\eqref{coy}. In order to show that the derivative vanishes, it is
sufficient to show that the three terms in
\eqref{mess}, besides the Lagrange
multipliers, are
independent of $a$, since then one can solve for the Lagrange
multipliers. Recall that
our ansatz for the solution has $y_{ai}=m_i/N$, independent of $a$.
The second term in \eqref{mess} is
then manifestly independent of $a$. Contrary to appearances,
the first term is also independent
of $a$, since $\sum_{b(\neq
a)}\wp(X_{ab})$ is, itself, independent of $a$ due to the form of
$X_a$ \eqref{bcase} and the elliptic periodicity of $\wp(z)$.
The second term involves
\EQ{
\sum_{b(\neq a)}\S^i_{ba}\S^j_{ab}\frac{\sigma(X_{ba}+z_{ji})}
{\sigma(X_{ba})\sigma(z_{ji})}\big(\zeta(X_{ba}+z_{ji})-\zeta(X_{ba})
\big)\ .
}
This is also independent of $a$. To show this, one uses the special
periodicity
property that we established for the quantity \eqref{blob}.
Hence, there exists Lagrange multipliers such that
$\partial H_0/\partial y_{ai}=0$.
The same kind of reasoning shows that $\partial H_j/\partial y_{ai}=0$.

Finally, the $t_{ai}$ derivatives of $H_i$ and $H_0$ involve
quantities like
\EQ{
\frac1{t_{ai}}\sum_{b(\neq a)}
\bigg\{\S^i_{ab}\S^j_{ba}\frac{\sigma(X_{ab}+z_{ji})}
{\sigma(X_{ab})\sigma(z_{ji})}f(X_{ab},z_{ji})
-\S^i_{ba}\S^j_{ab}\frac{\sigma(X_{ba}+z_{ji})}
{\sigma(X_{ba})\sigma(z_{ji})}f(X_{ba},z_{ji})\bigg\}\ ,
\label{mest}
}
where $f(X_{ab},z_{ji})$ is elliptic in $X_{ab}$. But this is of the
form \eqref{xad} which we have already shown
vanishes. \hspace*{\fill}{\it QED\/}

As in the softly broken $\N=4$ case, there are additional massive
vacua that are obtained by modular transformations in $\tau$
associated to the more general configurations \eqref{nvc}.

Finally, we can evaluate the Hamiltonians on the vacua. It suffices to
pick the vacua with $\rho_i=\lambda_i=1$ and $l=0$, since all the
others are related either by $\tau$ modular transformations or shifts
of the $\{z_i\}$. One finds
\SP{
H^*
\Big|_{l=0;\rho_i=\lambda_i=1}
&=-N\sum_im_i^2
\sum_{(r,s)\atop\neq(0,0)}
\wp\big(\tfrac{2r}q\omega_1+\tfrac{2s}p\omega_2\big)
+\sum_{i\neq j}m_im_j\Biggr\{
-N^2\tilde\wp(z_{ij})+N\wp(z_{ij})\\
&+
N^2\sum_{(r,s)\atop\neq(0,0)}\Big[\tilde\zeta(\Omega_{sr})-
\tfrac{2r}q\tilde\zeta(q\omega_2)-      
\tfrac{2s}p\tilde\zeta(p\omega_1)\Big]\Big[
\tilde\zeta(z_{ij}+\Omega_{sr})-\tilde\zeta(\Omega_{sr})\Big]\\
&+N\Big[(N-1)\tilde\zeta(z_{ij})-\sum_{(r,s)\atop\neq(0,0)}\big(
\tilde\zeta(z_{ij}+\Omega_{sr})-
\tilde\zeta(\Omega_{sr})\big)\Big]
\Big[\zeta(z_{ij})-\frac2k\sum_{l(\neq
i)}\zeta(z_{il})\Big]\\
&-(N-1)\sum_{(r,s)\atop\neq(0,0)}
\wp\big(\tfrac{2r}q\omega_1+\tfrac{2s}p\omega_2\big)\Biggr\}\ ,
\label{crazy1}
}
along with
\EQ{
H_i\Big\vert_{l=0;\rho_i=\lambda_i=1}=2N\sum_{j(\neq
i)}m_im_j\Big[(N-1)\tilde\zeta(z_{ij})-\sum_{(r,s)\atop\neq(0,0)}
\big(\tilde\zeta(z_{ij}+\Omega_{sr})-
\tilde\zeta(\Omega_{sr})\big)\Big]\ .
\label{crazy}
}

The expression for the residues, $H_i$, above matches \eqref{res}
precisely. It now remains to show that the expression for $H^*$ in
\eqref{crazy1} matches with the M-theory result \eqref{defhstar} using
the expression for 
$H_0$ in \eqref{constant}. This can be achieved by noting that both
the expressions for $H^*$ are in fact $\ttau$-elliptic in the
variables $z_{ij}$. It is then a straightforward but tedious excercise
(making use 
of various identities provided in Appendix A) to show
that the residues of the double poles and simple poles and the
constant pieces of these
expressions treated as functions of $z_{ij}$, are indeed identical. As
the expressions are $\ttau$-elliptic in the $z_{ij}$, this is sufficient to 
demonstrate that they are in fact the same.

Recall that in the M theory picture one has a freedom to choose which
of the $N$ 
branches of the
covering each NS5-brane lies on. This freedom, represented by the
parameters $(s_i,r_i)$ in \eqref{vas}, 
is encoded in the integrable system as the
$N$ choices for the roots of unity $\rho_i$ and $\lambda_i$. In
particular, the covering where all the NS5-branes lie on the same
branch, $(s_i,r_i)=(0,0)$ corresponds to $\rho_i=\lambda_i=1$.

\section{Conclusions and Future Directions}

In this paper we have provided a classification of the
vacuum structure and duality properties of the $\N=1^*$ deformations
(mass-deformations) of the $\N=2$ quiver theories. We have also
obtained exact results for certain chiral condensates in the massive
vacua of these theories, following two completely  different
approaches, namely: (i) by lifting the corresponding Type IIA brane
set-ups to M-theory and (ii) by studying the theory on ${\mathbb
R}^3\times S^1$. For a certain class of mass-deformations, both
approaches were used to independently evaluate the exact
superpotential (Eqs.\eqref{res},\eqref{constant} and
\eqref{specialsup}) in each massive vacuum and extremely non-trivial  
agreement was found. In particular, one of the main results of this
paper is an exact superpotential Eq. \eqref{spot} for the theory on ${\mathbb
R}^3\times S^1$, which coincides with a certain linear combination
of the quadratic Hamiltonians of the spin-generalization of the
elliptic Calogero-Moser system. This is a generalization of the
corresponding results for the mass-deformed $\N=4$ theory obtained in 
\cite{Nick}. Although we have only concentrated on the massive
vacua, the 
superpotential \eqref{spot} also determines all the massless
vacua. However, even in the basic $k=1$ case the structure of the
massless vacua is not known beyond $N=3$. It would be interesting to
understand the structure of these vacua as well.

Some immediate applications of our results include the calculation of
physical quantities such as the gluino condensate and tensions of
domain walls interpolating between the massive vacua
\cite{nickprem,inprogress}. In particular, these quantities may be
evaluated in the large $g^2N$, large-$N$ limit for a direct
comparison with the appropriate deformation of the Type IIB
backgrounds on $AdS_5\times S^5/{\mathbb Z}_k$. In the $k=1$ case
{\it i.e.\/} in the $N=1^*$ theory, the corresponding string backgrounds
\cite{polstr}
were asymptotically $AdS_5\times S^5$ containing D3-branes polarized
to 5-branes in the
interior, wrapping 2-cycles of the $S^5$. From the point of view of
the string dual, it would be interesting to understand the
characterisation of the extra vacua which arise in the $k>1$
theories. As remarked in \cite{polstr}, they are presumably associated
with the values of twisted sector fields. 

The superpotential Eq. \eqref{spot} and the expressions for the
condensates Eqs. \eqref{res}, \eqref{constant}, \eqref{defhstar} 
and \eqref{crazy1} contain a 
wealth of information regarding instanton and ``fractional instanton''
contributions in the massive vacua. In the three-dimensional picture, the
superpotential receives contributions from semiclassical monopoles
carrying fractional topological charge. The nature of these
contributions is visible in the condensates in the semiclassical
limit. For example in the $N^{k-1}$ confining vacua with $p=N, q=1$
and $l=0$, the semi-classical expansion reveals contributions from
instantons as well as fractional instantons in each gauge group
factor. In the $k=1$ theory, {\it i.e.\/} mass-deformed $\N=4$ SUSY
Yang-Mills, in the large $g^2N$ limit, the fractional instanton series
can be `resummed' using $\tilde S$-duality to obtain a dual
expansion. Terms in this dual expansion can be elegantly described in
the IIB string dual of Polchinski and Strassler as arising from world-sheet
instantons wrapping the 2-cycles of the 5-branes polarized from the D3-branes.
While one expects a similar interpretation to arise in the $k>1$
theories, the appearance of different types of fractional
instantons (from each gauge group factor) and their associated
actions needs to be understood better from the point-of-view of the
string dual. 

Finally, we point out that in the mass-deformed $\N=4$ theory on
${\mathbb R}^3\times S^1$, the appearance of the elliptic
superpotential \cite{Nick} encoding
pairwise interactions between $N$ particles on a torus can be given a very
nice, physical interpretation. Realizing the compactified $\N=4$ theory on
$N$ D3-branes wrapped on a circle, one may perform T-duality and lift
the resulting setup of $N$ D2-branes to M-theory. We thus obtain $N$
M2-branes with two transverse compact directions (the second compact
direction being the M-dimension), which may now be thought of as the
$N$ particles on a torus. Upon introducing the $\N=1^*$
mass-deformation the M2-branes exert forces on each other which is
described by the Weierstrass superpotential obtained in
\cite{Nick}. It would be extremely interesting to generalize this 
picture to the superpotential Eq.\eqref{spot} for the $\N=1^*$ quiver
theory and obtain an interpretation for the ``spin-spin'' interactions
in the Hamiltonian in terms of appropriate interactions between
M2-branes.  

\startappendix

\Appendix{Some Properties of Elliptic Functions}

In this appendix we provide some useful---but far from
complete---details of elliptic functions and their near cousins.
For a more complete
treatment we refer the reader to one of the textbooks, for example
\cite{WW}. An elliptic function $f(z)$ is a function on the complex
plane, periodic in two periods $2\omega_1$ and $2\omega_2$. We will
define the lattice $\Gamma=2\omega_1{\mathbb Z}\oplus2\omega_2{\mathbb
Z}$ and define the basic period parallelogram as
\EQ{
{\cal
D}=\big\{z=2\mu\omega_1+2\nu\omega_2,\ 0\leq\mu<1,\ 0\leq\nu<1\big\}\ .
}
The archetypal
elliptic function is the Weierstrass $\wp(z)$ function. It is an even
function which can be defined via
\EQ{
\wp(z)=\frac1{z^2}+\sum_{m,n\atop\neq(0,0)}
\Big\{\frac1{(z-2m\omega_1-2n\omega_2)^2}
-\frac1{(2m\omega_1+2n\omega_2)^2}\Big\}\ ,
}
where the sums is over all integer pairs except $m=n=0$. The function
$\wp(z)$ is analytic throughout ${\cal D}$, except at $z=0$
where it has a double pole:
\EQ{
\wp(z)=\frac1{z^2}+{\cal O}(z^2)\ .
}

The other two important functions for our purposes, are $\sigma(z)$ and
$\zeta(z)$. They are both odd functions but are {\it not\/} elliptic,
since they have anomalous
transformations under shifts by periods:
\EQ{
\zeta(z+2\omega_\ell)=\zeta(z)+2\zeta(\omega_\ell)\ ,\qquad
\sigma(z+2\omega_\ell)=-\sigma(z)e^{2(z+\omega_\ell)\zeta(\omega_\ell)}\
.
}
The three functions are related via
\EQ{
\zeta(z)=\frac{\sigma'(z)}{\sigma(z)}\ ,\qquad\wp(z)=-\zeta'(z)\ .
}
In ${\cal D}$,
$\zeta(z)$ has a simple pole and $\sigma(z)$ a simple zero at $z=0$:
\EQ{
\zeta(z)=\frac1z+{\cal O}(z^3)\ ,\qquad\sigma(z)=z+{\cal O}(z^5)\ .
}
Some useful identities for $\zeta(z)$ and $\wp(z)$ evaluated on
half-periods are
\AL{
&\omega_2\zeta(\omega_1)-\omega_1\zeta(\omega_2)=\frac{\pi i}2\ ,\\
&\zeta(\omega_1+\omega_2)=\zeta(\omega_1)+\zeta(\omega_2)\ ,\\
&\wp(\omega_1+\omega_2)+\wp(\omega_1)+\wp(\omega_2)=0\ .
}

An elliptic function $f(z)$ can always be expressed as
\EQ{
f(z)=c_0+\sum_{k=1}^n\big\{c_{k,1}\zeta(z-a_k)+\cdots+
c_{k,r_k}\zeta^{(r_k-1)}(z-a_k)\big\}
}
for constants $c_0$ and $c_{k,i}$ with the restriction that the sum of
the simple pole residues vanishes, $\sum_{k=1}^nc_{1,k}=0$.
In particular, an elliptic function which has no poles is a constant.

Of particular importance to us is the behaviour of our basic functions
under modular transformations of the complex structure of the
torus defined by $\Gamma$.
These $\SL(2,{\mathbb Z})$ transformations act as
\EQ{
\MAT{\omega_2\\ \omega_1}\to\MAT{a&b\\ c&d}\MAT{\omega_2\\ \omega_1}\
,
\label{vv}
}
for $a,b,c,d\in{\mathbb Z}$ subject to $ad-bc=1$. Since we choose
$\omega_1=i\pi$ and $\omega_2=i\pi\tau$,
after the transformation \eqref{vv}, one has to
perform a re-scaling by $(c\tau+d)^{-1}$ so that the transformation on
$\tau$ has the usual form:
\EQ{
\tau\to\tau'=\frac{a\tau+b}{c\tau+d}\ .
\label{tt}
}
A function $f(z)$ has modular weight $w$ if
\EQ{
f(z|\tau')=(c\tau+d)^wf(z(c\tau+d)|\tau)\ .
}
The functions $\wp(z)$, $\zeta(z)$ and $\sigma(z)$ have modular
weights 2, 1 and $-1$, respectively.

We will need to explore the semi-classical limit of our functions. This is the
limit $g^2\to0$, $\tau\to i\infty$ or $\omega_2\to-\infty$ with fixed
$\omega_1=i\pi$. In this limit, for $z\in{\cal D}$,
\SP{
&\wp(z)\to \frac1{12}+\frac1{4\sinh^2\tfrac z2}\ ,\\
&\zeta(z)\to -\frac z{12}+\tfrac12{\rm coth}\tfrac
z2\ ,\\
&\sigma(z)\to 2e^{-z^2/24}\sinh\tfrac z2\ .
}
In particular, for fixed $-2<\alpha_i<2$, $\sum_i\alpha_i=0$,
\EQ{
\lim_{\tau\to i\infty}\sum_i\zeta(\alpha_i\omega_2)=
-\tfrac12\sum_i{\rm sign}(\alpha_i)\ .
}

Below we collect, without proof,
various useful identities. The
un-tilded functions are defined with respect to the lattice
$\Gamma=2\omega_1{\mathbb Z}\oplus2\omega_2{\mathbb Z}$ while the
tilded functions are defined with respect to the lattice
$\tilde\Gamma=2p\omega_1{\mathbb Z}\oplus2q\omega_2{\mathbb Z}$.
\EQ{
\zeta(z)=\sum_{(r,s)}\tilde\zeta(z+2s\omega_1+2r\omega_2)-
\sum_{(r,s)\atop\neq(0,0)}\tilde\zeta(2s\omega_1+2r\omega_2)-
\frac{q\tilde\zeta(p\omega_1)-\zeta(\omega_1)}{\omega_1}z\ .
}
\EQ{
\wp(z)=\sum_{(r,s)}\tilde\wp(z+2s\omega_1+2r\omega_2)-
\sum_{(r,s)\atop\neq(0,0)}\tilde\wp(2s\omega_1+2r\omega_2)\ .
}
\EQ{
\sum_{(r,s)\atop\neq(0,0)}
\wp\big(\tfrac{2r}q\omega_1+\tfrac{2s}p\omega_2\big)
=-pq\sum_{(r,s)\atop\neq(0,0)}
\tilde{\wp}\big(2s\omega_1+2r\omega_2\big)\ .
\label{convert}}
\EQ{
\sum_{(r,s)\atop\neq(0,0)}
\Big\{\zeta\big(\tfrac{2r}q\omega_1+\tfrac{2s}p\omega_2\big)-
\tfrac{2r}q\zeta(\omega_1)-\tfrac{2s}p\zeta(\omega_2)\Big\}^2=
\frac{pq-2}{pq}\sum_{(r,s)\atop\neq(0,0)}
\wp\big(\tfrac{2r}q\omega_1+\tfrac{2s}p\omega_2\big)\ .
}
\SP{
&\sum_{(r,s)\atop\neq(0,0)}
\Big\{\zeta\big(\tfrac{2r}q\omega_1+\tfrac{2s}p\omega_2\big)-
\tfrac{2r}q\zeta(\omega_1)-\tfrac{2s}p\zeta(\omega_2)\Big\}e^{-2\pi
i\tfrac{rr'}q+2\pi i\tfrac{ss'}p}\\
&\qquad\qquad
=pq\Big\{\tilde\zeta\big(2s'\omega_1+2r'\omega_2\big)-
\tfrac{2r'}q\tilde\zeta(q\omega_2)-\tfrac{2s'}p\tilde\zeta(p\omega_1)\Big\}\ .
}
\EQ{
\Big[\zeta(x+y)-\zeta(x)-\zeta(y)\Big]^2=\wp(x+y)+\wp(x)+\wp(y)\ .
}
For $\sum_i\lambda_i=0$
\SP{
\Big(\sum_i\lambda_i\zeta(z_i)\Big)^2&=\sum_i\lambda_i^2\wp(z_i)\\
&+\sum_{i\neq
j}\lambda_i\lambda_j\Big\{2\zeta(z_i)\zeta(z_j-z_i)-\tfrac12
\wp(z_i-z_j)+\tfrac12\zeta(z_i-z_j)^2\Big\}\ .
\label{imp}}
Finally
\EQ{
\sum_{(r,s)\atop\neq(0,0)}
\wp\big(\tfrac{2r}q\omega_1+\tfrac{2s}p\omega_2\big)=
{pq\over 12}\left[E_2(\tau)-{q\over p}E_2(\ttau)\right]\ ,
\label{e2p}
}
where $E_2(\tau)$ is the second Eisenstein series \cite{koblitz} which
has the modular 
transformation properties $E_2(\tau+1)=E_2(\tau)$ and 
$E_2(\tau)=E_2(-1/\tau)/\tau^2-6/i\pi\tau$.

\Appendix{Rotating the brane configuration}

For the non-elliptic models it is well established that the breaking
supersymmetry by adding mass terms for the adjoint chiral multiplets
can be realized in the Type IIA brane configurations by relative rotations of
the NS5-branes into the $w=x^8+ix^9$ direction \cite{Hori:1998ab}.
This way of realizing
soft breaking to $\N=1$ in the non-elliptic models gives a very simple
picture of why the Riemann surface $\Sigma$ has to degenerate at a
vacuum. The point is that only very particular surfaces $\Sigma$ are
``rotatable''. Let us quickly review the argument in the context of a
non-elliptic model with two NS5-branes describing a model with $\U(N)$
gauge symmetry. Breaking to $\N=1$ is achieved by rotating the first
NS5-brane into the $(v,w)$ plane, along the line $w=\mu v$,
while leaving the second NS5-brane intact. Since $v$ diverges at the
positions of the five-branes, $w$ must diverge at the first five-brane
like $\mu v$, but vanish at the position of the second five-brane.
Since $\mu$ is a
free parameter we can construct the rotated surface $\widetilde\Sigma$
order-by-order in $\mu$. Working to first order in $\mu$ allows us to
find the constraints on the original $\Sigma$ in order that it be
``rotatable''. To first order, the problem is to find a meromorphic
function $w$
on the initial surface $\Sigma$, which has a simple pole at the
position of the first five-brane.\footnote{Technically $\Sigma$ is
non-compact at the positions of the five-brane, so we must compactify
it by adding these points to get a compact surface.}
But a function with a single simple
pole cannot exist on surfaces of genus $>0$, and so in order that $\Sigma$ be
rotatable it must have completely degenerated to a surface of genus 0.

The goal of the present section will be to generalize
the brane rotation story to the elliptic models. We shall see that the
situation is rather more involved but the essence of the problem is
the same. The condition that a surface $\Sigma$ be rotatable will
boil down to the existence, or otherwise, a certain meromorphic
function on $\Sigma$ with a prescribed pole structure.
To start with we consider the original $\N=1^*$ models where there is
only a single NS5-brane. The obvious problem is that there is
no immediate sense rotating a single NS5-brane. Thinking about the
periodicity in terms of 
an infinite string of images, what we need to do is rotate
each image relative to the last. The situation is rather
similar to the introduction of the global $\N=2$ mass $m$. Recall that
in order to introduce this mass we had to modify the spacetime itself
introducing a non-trivial bundle over $E_\tau$.
We need to do an analogous thing in order to
incorporate the $\N=1^*$ rotation. The twist acts as a complex
rotation on the complex combination
$g=v+iw$:
\EQ{
g\to e^{i\xi} g\ ,\qquad e^{i\xi}\equiv\sqrt{\frac{1+i\mu}{1-i\mu}}\ ,
}
where $\mu$ is the $\N=1^*$ mass. Comparing with the discussion of how
the global mass $m$ was introduced, we can trivialize the
bundle at the
expense introducing a suitable singularity at an arbitrary point of
$E_\tau$. Recalling that the shift $v\to v+m$ required that
$v(z)$ to a simple pole
at an arbitrary point (chosen to be $z=0$) on each sheet with residue
$-m/N$, we see that $g(z)$ must have an essential singularity of the form
\EQ{
g(z)\thicksim \exp\Big(-\frac{i\xi/N}{z-z_0}\Big)
\label{ess}
}
at a new arbitrary point $z_0$ on each sheet. The function $g(z)$ must
also have simple poles at $z=0$, as before to incorporate the global
mass $m$, on each sheet and also a simple pole of residue $m$ at
$z=z_1$ on the single sheet where the NS5-brane is located.

We now follow the logic of
the non-elliptic case and work to first order in $\mu$ in order to
derive the condition that $\Sigma$ is rotatable. To leading order we
think of $g(z)$
as a function on $\Sigma$. When we turn off the rotation
$g(z)\equiv v^{\sst(0)}(z)$, where $v^{\sst(0)}(z)$ describes the
un-rotated surface $\Sigma$. Consider the function
$h(z)=g(z)-v^{\sst(0)}(z)$. The simple pole at the NS5-brane is now
cancelled and so $h(z)$ has simple poles at $z=0$ on each sheet, of
the same residue, and essential singularities at $z=z_0$ of the form
\eqref{ess} on each sheet. It is instructive to consider the function
$\sum_ah_a(z)$,
a sum over the images of $h(z)$ on each sheet. This is a {\it bona
fide\/} meromorphic function on the torus $E_\tau$ itself with a
simple pole at $z=0$ and an essential singularity of the form
\eqref{ess} at $z=z_0$. The unique function with these properties is
\EQ{
\frac{\sigma(z+i\xi/N)}{\sigma(z)}e^{-i\xi\zeta(z-z_0)/N}\ .
\label{haas}
}
We can find a necessary condition that $\Sigma$ be rotatable by
working to first order in $\mu$ (or $\xi$). To this order,
we expand \eqref{ess} to find simple poles at $z=z_0$, with residue
$-i\mu/N$ on each sheet. So the surface
$\Sigma$ will be rotatable provided there exists a meromorphic
function $h(z)$ on $\Sigma$, more properly its compactification,
with simple poles on each sheet at $z=0$ and $z_0$ of equal and
opposite residue $\pm
i\mu/N$, respectively.\footnote{This last
requirement follows from the fact that $\sum_ah_a(z)$
is a meromorphic function on the torus
$E_\tau$ with two simple poles the sum of whose residues must vanish,
as can be verified by expanding \eqref{haas} to linear order in $\xi$.}

We can translate the condition into something more convenient by the
following chain of arguments. Since $z_0$ is arbitrary, we
can take $z_0\to0$. In this limit the simple poles
merge together
in pairs on each sheet yielding double poles. So
$\Sigma$ will be rotatable provided there exists a meromorphic
function on $\Sigma$ with double poles at $z=0$ on each sheet with the
same coefficient and vanishing residue.
Finally, by taking a linear combination of this
function, $v^{\sst(0)}(z)^2$ and $v^{\sst(0)}(z)$ we can find a
function which is now regular at $z=0$, but now has a double pole at
$z=z_1$ on the single sheet where the NS5-brane is located. So a
necessary condition that $\Sigma$ be rotatable is that there exists a
meromorphic function $f(z)$ on its compactification which has a single double
pole at the position of the NS5-brane.

The Riemann-Roch Theorem implies that, generically, $f(z)$ can only
exist on surfaces with genus $0$ or $1$. So generically in our
example, in order for
$\Sigma$ to be rotatable it must
completely degenerate to an unbranched (unramified) $N$-fold cover of
the torus. The function with a single double pole is then the
Weierstrass function. In this case, as we shall see, the theory has a mass gap
(ignoring the overall $\U(1)$ factor).
Exceptionally, however, a suitable function $f(z)$
can exist on a higher genus surface, in which case the deformed theory
is massless since there is an unbroken $\U(1)^{g-1}$ symmetry (in
addition to the overall $\U(1)$ factor) where $g$ is the genus.
In fact, we can say more
about these exceptional surfaces. In order that there exists a
meromorphic function with a single double pole at the position of the
NS5-brane $\Sigma$ must necessarily be {\it hyper-elliptic\/} and
moreover the NS5-brane must necessarily
be positioned at one of the $2g+2$ Weierstrass Points of the surface.

The generalization to the case with more NS5-branes is now clear. As
well as an overall rotation, there are $k-1$
relative rotations of the NS5-branes parameterized by
$\hat\mu_i=\mu_i-\mu/k$,
$i=1,\ldots,k$ and $\sum_i\mu_i=\mu$. The surface will be rotatable
if $h(z)$ has the pole structure established in the $k=1$ case above
but, in addition, has a
simple pole at the position of the $i^{\rm th}$ NS5-brane with residue
$\hat\mu_i$ for $i=1,\ldots,k$.
For example, if $\mu=0$, then $h(z)$ only has
simple poles at the NS5-branes with residues $\hat\mu_i$.
As before the Riemann-Roch Theorem dictates that
such a function will generically only exist when $\Sigma$
undergoes complete degeneration to the torus. However, exceptionally
there will exist higher genus surfaces which admit such a function. Notice
that these exceptional $g>1$
surfaces will depend on the $\mu_i$: in other
words, as we vary the $\N=1$ deformation the rotatable
surface $\Sigma$ will change accordingly.

\end{document}